\documentclass[a4paper,fleqn,usenatbib]{mnras}

\usepackage{newtxtext,newtxmath}

\usepackage[T1]{fontenc}
\usepackage{ae,aecompl}

\usepackage{graphicx}	
\usepackage{amsmath}	
\usepackage{amssymb}	
\usepackage{pdflscape}
\usepackage[margin=5pt]{subfig}

\title[Rosetta spacecraft potential at comet 67P]{Measurements of the electrostatic potential of Rosetta at comet 67P}

\author[E. Odelstad et al.]{
Elias Odelstad,$^{1,2}$\thanks{E-mail: elias.odelstad@irfu.se}
G. Stenberg-Wieser,$^{3}$
M. Wieser,$^{3}$
A. I. Eriksson,$^{2}$
H. Nilsson,$^{3}$
\newauthor and F. L. Johansson$^{1,2}$
\\
$^{1}$Department of Physics and Astronomy, Uppsala University, Box 516, 75120 Uppsala, Sweden\\
$^{2}$Swedish Institute of Space Physics, Box 537, 75121 Uppsala, Sweden\\
$^{3}$Swedish Institute of Space Physics, Box 812, 98128 Kiruna, Sweden
}

\date{Accepted XXX. Received YYY; in original form ZZZ}

\pubyear{2017}

\begin{document}
\label{firstpage}
\pagerange{\pageref{firstpage}--\pageref{lastpage}}
\maketitle

\begin{abstract}
We present and compare measurements of the spacecraft potential ($V_{\textrm{s/c}}$) of the Rosetta spacecraft throughout its stay in the inner coma of comet 67P/Churyumov-Gerasimenko, by the Langmuir probe (RPC-LAP) and Ion Composition Analyzer (RPC-ICA) instruments. $V_{\textrm{s/c}}$ has mainly been negative, driven by the high temperature ($\sim$5-10~eV) of the coma photoelectrons. The magnitude of the negative $V_{\textrm{s/c}}$ traces heliocentric, cometocentric, seasonal and diurnal variations in cometary outgassing, consistent with production at or inside the cometocentric distance of the spacecraft being the dominant source of the observed plasma. LAP only picks up a portion of the full $V_{\textrm{s/c}}$ since the two probes, mounted on booms of 2.2 and 1.6 m length, respectively, are generally inside the potential field of the spacecraft. Comparing with the minimum energy of positive ions collected by ICA, we find numerous cases with strong correlation between the two instruments, from which the fraction of $V_{\textrm{s/c}}$ picked up by LAP is found to vary between about 0.7 and 1. We also find an ICA energy offset of 13.7 eV (95\% CI: [12.5, 15.0]). Many cases of poor correlation between the instruments are also observed, predominantly when local ion production is weak and accelerated ions dominate the flux, or during quiet periods with low dynamic range in $V_{\textrm{s/c}}$ and consequently low signal-to-noise ratios.
\end{abstract}

\begin{keywords}
plasmas -- comets: individual: 67P -- methods: data analysis -- methods: statistical -- space vehicles: instruments -- instrumentation: detectors
\end{keywords}



\section{Introduction}

The European Space Agency's Rosetta spacecraft arrived at comet 67P/Churyumov-Gerasimenko in August 2014 and was immersed in the cometary plasma until the end of September 2016. At arrival, the heliocentric distance was 3.5 AU, decreasing to 1.26 AU at perihelion one year later and then increasing again, reaching 3.83 AU at the end of mission (EOM). The instruments of the Rosetta Plasma Consortium (RPC) \citep{Carr2007} have thus been able to follow the evolution of the cometary plasma environment and the varying spacecraft-plasma interaction over an approximately four orders of magnitude variation in plasma density, driven by an order of magnitude change of solar illumination from rendez-vous to perihelion and varying distance of Rosetta to the comet nucleus.

\subsection{Ions}

The first cometary plasma to be detected was cometary pick-up ions at a distance of 100 km from the nucleus on August 7, 2014, by RPC-ICA \citep{Nilsson2015}. These were water ions at nearly 100 eV, created upstream and accelerated towards the spacecraft by the convective electric field perpendicular to the solar wind direction. The first locally produced ions were detected by RPC-IES on August 19, 2014, at a distance of $\sim$80 km from the nucleus \citep{Goldstein2015}. These were also seen by ICA from September 21, 2015, at a distance of 28 km, and had typical energies of 5-10 eV, close to the negative of the spacecraft potential. It is unclear whether the appearance of these ions in ICA was triggered by their local density increasing above the measurement threshold of the instrument, or if it was because the increasingly negative spacecraft \citep{Yang2016} started pulling them in over the instrument energy threshold. Possibly, it is a combination of both effects.

Deflection of the solar wind ions was also first observed around September 21, 2014, with protons being deflected by about $20^\circ$ \citep{Nilsson2015}. The total plasma density was typically on the order of $5 - 10$ cm$^{-3}$ in September 2014, at $\sim$30 km from the nucleus. This is comparable to the solar wind proton density, but the mass density is about an order of magnitude larger.  In addition, detection of He$^+$ ions showed that charge exchange reactions had begun to occur, since these ions are created by charge exchange between solar wind He$^{2+}$ and cometary water molecules. Thus, the solar wind was already clearly influenced by interaction with the cometary plasma. The deflection angle of solar wind protons increased dramatically over the following months, peaking at values between 140$^\circ$ and 180$^\circ$ (i.e.\ sunward) between March and May 2015. After that, at a heliocentric distance of at 1.76~AU, the solar wind (both protons and alpha particles) was lost completely and did not reappear again until December 2015, after perihelion at a heliocentric distance of 1.64~AU \citep{Behar2017}.

The flux of accelerated cometary water ions increased dramatically between August 2014, at 3.6 AU, and March, 2015, at 2.0 AU, on average by 4 orders of magnitude \citep{Nilsson2015b}. This was observed also further away from the nucleus, during the excursions out to 250 km from the nucleus in February, 2015. The flux of these ions was subsequently relatively constant due to Rosetta's orbit changing with comet activity, although there was a slight decrease of the flux, and the maximum energy of the ions, around perihelion. Thus, these cometary pick-up ions were observed even after the solar wind signature was lost from the instrument \citep{Nilsson2017}.

\subsection{Electrons and spacecraft potential}

\citet{Odelstad2015} analysed RPC-LAP spacecraft potential measurements from the beginning of September 2014 to the end of March 2015, finding it to be generally negative (often by several tens of volts). This was attributed to the bulk electron temperature being quite high, $\sim 5-10$ eV, a consequence of the low collision rate in the tenuous neutral gas of the inner coma. The plasma ions and weak spacecraft photoemission could not balance the flux of such warm electrons to the spacecraft, which thus became negatively charged. This greatly affected the particle and dust measurements by instruments on the spacecraft main body, e.g.\ effectively shielding the spacecraft from small negatively charged dust grains \citep{fulle2015}.

In addition to these warm thermal electrons, a supra\-thermal electron population, accelerated up to several hundreds of eV, was detected by IES \citep{Clark2015}. Their origin is still unclear, but they appear to become more numerous during periods of stormy solar wind \citep{Edberg2016}, which might indicate that the responsible heating mechanism is connected to the solar wind energy input. \citet{Broiles2016} showed that this supra\-thermal population actually consists of multiple sub-populations and suggested lower hybrid waves as a possible acceleration mechanism. Such waves have since been identified in the Rosetta data set \citep{Karlsson2017, Andre2017}, although their correlation with electron heating remains to be investigated. There was a general trend of increasing fluxes of these supra\-thermal electrons during the first months of the mission, resembling the increase in accelerated water ion flux observed by ICA.

Finally, a third population of cold electrons, with characteristic energies of less than 0.1 eV, has been identified in the data from LAP \citep{Eriksson2017a}. These were observed very intermittently as pulses typically lasting for a few seconds to a few tens of seconds as seen in the spacecraft frame. They presumably obtain their low temperatures from cooling by collisions with neutrals in the densest inner part of the coma, from which filaments seem to detach and be transported outward \citep{Koenders2015, Henri2017}.

\subsection{Morphology and evolution of the cometary neutral gas and plasma}

The tilted rotation and complex geometry of the nucleus \citep{Sierks2015} produced strong diurnal and seasonal variations in the comet outgassing. Up until southward equinox in May 2015, the northern (w.r.t.\ the rotation axis) hemisphere of the nucleus was tilted towards the sun, with most of the gas and dust coming from this (summer) hemisphere and with the neck region between the two lobes being the most active part \citep{Hassig2015, Sierks2015, Gulkis2015, Bockelee2015}. This came through also in the near-nucleus ($\lesssim 50$ km) plasma environment, where the plasma density \citep{Edberg2015} and spacecraft potential \citep{Odelstad2015}, which traces the thermal flux of warm electrons in the cometary plasma (cf.\ Section \ref{sec:Vsc_monitor}), peaked over the neck region in the northern hemisphere, closely following the neutral density.

\citet{Edberg2015} reported densities of a few tens of cm$^{-3}$ over the lobes up to about 300 cm$^{-3}$ over the neck region in the northern hemisphere in October 2014, when at 10 km from the nucleus. The total neutral density was found to fall off as $1/r^2$ with distance $r$ from the nucleus \citep{Hassig2015} while the plasma density decayed as $1/r$, consistent with a plasma produced at or inside the position of the spacecraft and expanding radially outward at constant speed. However, this interpretation requires the absence of any significant solar wind electric field. Possibly, this field is quenched close to the nucleus by significant ion pickup and mass loading, as indicated by the solar wind deflection observed by ICA. At perihelion, plasma densities reached several thousands cm$^{-3}$ \citep{Vigren2017, Henri2017}.

The peak water production shifted from northern to southern latitudes around the southward equinox in May 2015 \citep{Hansen2016}. The transition period was characterised by a complex water distribution, driven by rotation and active areas in the north and south and generally a minimum in or near the equatorial plane. The seasonal dependence of the water production became more clearcut further into southern summer, with clear peaks above the southern hemisphere. Maximum water production was attained about 20 days after perihelion. This will be compared to the evolution of the spacecraft potential in Section \ref{sec:evolution}.

\subsection{Objective and scope of this study}

This Paper presents and compares two methods for obtaining spacecraft potential measurements by RPC-LAP and evaluates them by a comprehensive comparison to RPC-ICA ion measurements, with the aim of improving the analysis and interpretation of these measurements by means of a cross-calibration. We also study the evolution of the spacecraft potential throughout the mission and discuss what this implies for the cometary plasma environment, essentially a continuation and expansion of the analysis by \citet{Odelstad2015}.

\section{Instrumentation and measurements}

\subsection{Spacecraft potential measurements by RPC-LAP}

\subsubsection{The RPC-LAP instrument}

RPC-LAP consists of two spherical Langmuir probes (LAP1 and LAP2) with radii of 2.5 cm, mounted on 15 cm stubs on the tips of booms of length 2.24 m and 1.62 m, respectively. LAP1, being mounted on the longer boom at an angle of 45$^\circ$ off nominal nadir, is positioned to minimize the disturbances from the spacecraft sheath and wake effects without violating the field of view of other instruments \citep{Eriksson2007}. In this Paper, we include only measurements from LAP1.

LAP is a very versatile instrument, in principle capable of obtaining the electron density and temperature, ion density and flow speed, spacecraft potential, mean ion mass and integrated UV flux. However, the highly variable and evolving plasma environment of comet 67P makes many of these measurements difficult when it comes to consistent automatic analysis covering longer time periods \citep{Eriksson2017a}. The spacecraft potential measurements stand out in this regard by providing consistent and reliable results during the bulk of the mission. However, the LAP spacecraft potential measurements only represent a portion of the full spacecraft potential ($V_{\textrm{s/c}}$) due to the fact that the two probes are generally inside the potential field of the spacecraft \citep{johansson2016sctc}. When local production is strong, which is likely most of the time, information about $V_{\textrm{s/c}}$ can also be derived from the collection of low energy ions by the Ion Composition Analyzer (RPC-ICA) \citep{Nilsson2007,Nilsson2015}, located on the main spacecraft body, using the minimum energy of collected positive ions. Comparison of the two gives information on the electrostatic potential around Rosetta and allows determination of the fraction of $V_{\textrm{s/c}}$ observed by LAP.

\subsubsection{$V_{\textrm{s/c}}$ as a monitor of the plasma environment}
\label{sec:Vsc_monitor}

Neglecting the typically much weaker ion currents (c.f.\ Sections \ref{sec:ion_current} and \ref{sec:Ii_SC}), the current exchange between the spacecraft and the plasma is dominated by impacting plasma electrons and photoemission of electrons from sunlit parts of the spacecraft surface. For a negatively charged spacecraft the current due to impacting plasma electrons is given by

\begin{equation}
	I_\textrm{e} = \underbrace{A_\textrm{s/c}}_{\textrm{S/C area}} \cdot \underbrace{n_\textrm{e} \sqrt{\frac{k_{\textrm{B}} T_\textrm{e}}{2\pi m_\textrm{e}}}}_{\mathcal{F}_\textrm{e}\textrm{, random e$^-$ flux}} \cdot e \cdot\underbrace{\exp\left\{\frac{e V_\textrm{s/c}}{k_{\textrm{B}} T_\textrm{e}}\right\}}_{\textrm{repelling factor}},
\label{eq:Ie}
\end{equation}
where $A_{s/c}$ is the total current-collecting area of the spacecraft and $n_\textrm{e}$, $T_\textrm{e}$, $m_\textrm{e}$ and $e$ are the electron number density, temperature, mass and charge, respectively, and $k_{\textrm{B}}$ is Boltzmann's constant. The photoemission current $I_\textrm{ph}^\textrm{s/c}$ from a negative spacecraft is independent of $V_{\textrm{s/c}}$, varying only with the heliocentric distance and solar EUV flux. For the LAP1, the photoemission throughout the mission is presented by \citet{Johansson2017}. Equating $I_\textrm{e}$ and $I_\textrm{ph}^\textrm{s/c}$ and solving for $V_{\textrm{s/c}}$ in Equation (\ref{eq:Ie}) gives

\begin{equation}
	V_{\textrm{s/c}} = -\frac{k_{\textrm{B}} T_\textrm{e}}{e} \log \left\{ \frac{A_\textrm{s/c} n_\textrm{e} e}{I_\textrm{ph}^\textrm{s/c}} \sqrt{\frac{k_{\textrm{B}} T_\textrm{e}}{2\pi m_\textrm{e}}}\right\}
\label{eq:Vsc}.
\end{equation}
Thus $V_{\textrm{s/c}}$ is essentially proportional to the logarithm of the random thermal flux of electrons $\mathcal{F}_\textrm{e}$ in the ambient plasma:

\begin{equation}
	V_\textrm{s/c} \propto -T_\textrm{e} \log \mathcal{F}_\textrm{e} \quad.
\label{eq:Vscprop}
\end{equation}
This allows $V_{\textrm{s/c}}$ to be used effectively as a long-term monitor of the cometary plasma environment \citep{Odelstad2015}.

Equation (\ref{eq:Vsc}) has been used together with photoemission current densities from LAP1 and assumed electron temperatures to estimate electron densities in \citet{Galand2016} and  \citet{Heritier2017a} and together with densities from the Rosetta Mutual Impedance Probe (RPC-MIP) to estimate electron temperatures \citep{Harja2017}.

\subsubsection{$V_{\textrm{s/c}}$ from Langmuir probe bias potential sweeps}
\label{sec:Vsc_sweeps}

\begin{figure}
	\includegraphics[width=\columnwidth]{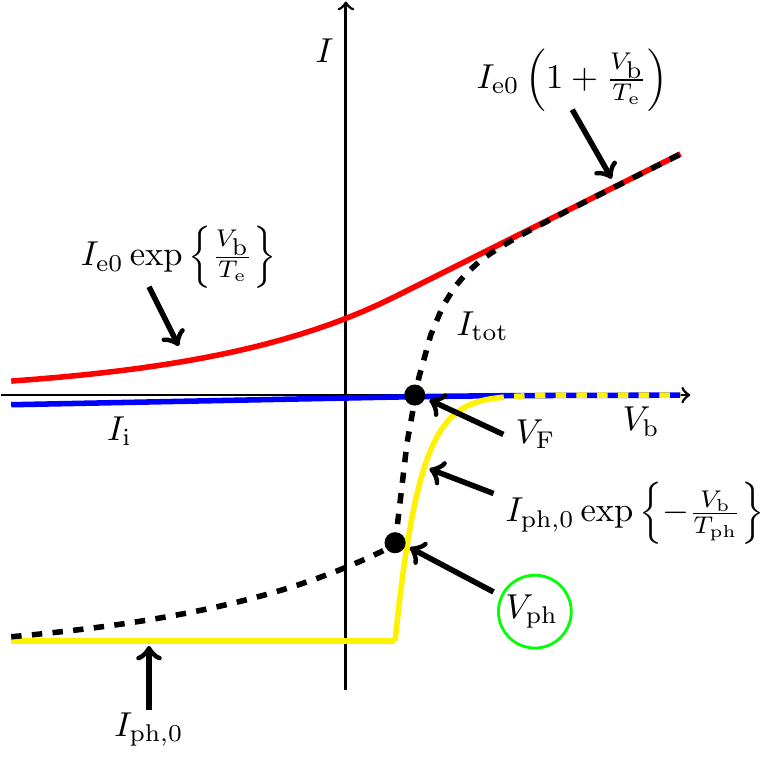}
	\caption{Illustration of the current-voltage relationship of a sunlit spherical Langmuir probe (dashed black line) in a warm tenuous plasma. The current components due to collection of ambient plasma electrons $I_\textrm{e}$ (red line), ions $I_\textrm{i}$ (blue line) and photoemission of electrons from the probe surface $I_\textrm{ph}$ (yellow line) are also shown. Here, $I_\textrm{e0} = A_\textrm{p} \mathcal{F}_\textrm{e}$ is the random thermal electron current to an uncharged probe with surface area $A_\textrm{p}$, $I_\textrm{ph,0}$ is the photo-saturation current and $T_\textrm{e}$ and $T_\textrm{ph}$ are the ambient electron and photoelectron temperatures in eV, respectively.}
	\label{fig:IVcurve}
\end{figure}

LAP $V_{\textrm{s/c}}$ measurements can be obtained from Langmuir probe bias potential sweeps, where the probe bias potential $V_\textrm{b}$ is sequentially stepped through a range of values, the maximum range being from -30~V to +30~V, and the probe current sampled at each value of $V_\textrm{b}$. The probe current is the sum of current contributions from ions and electrons from the ambient plasma and photoemission of electrons from the probe surface. The resulting current-voltage relationship, and the constituent currents, is illustrated in Figure \ref{fig:IVcurve}.

\begin{figure}
	\includegraphics[width=\columnwidth]{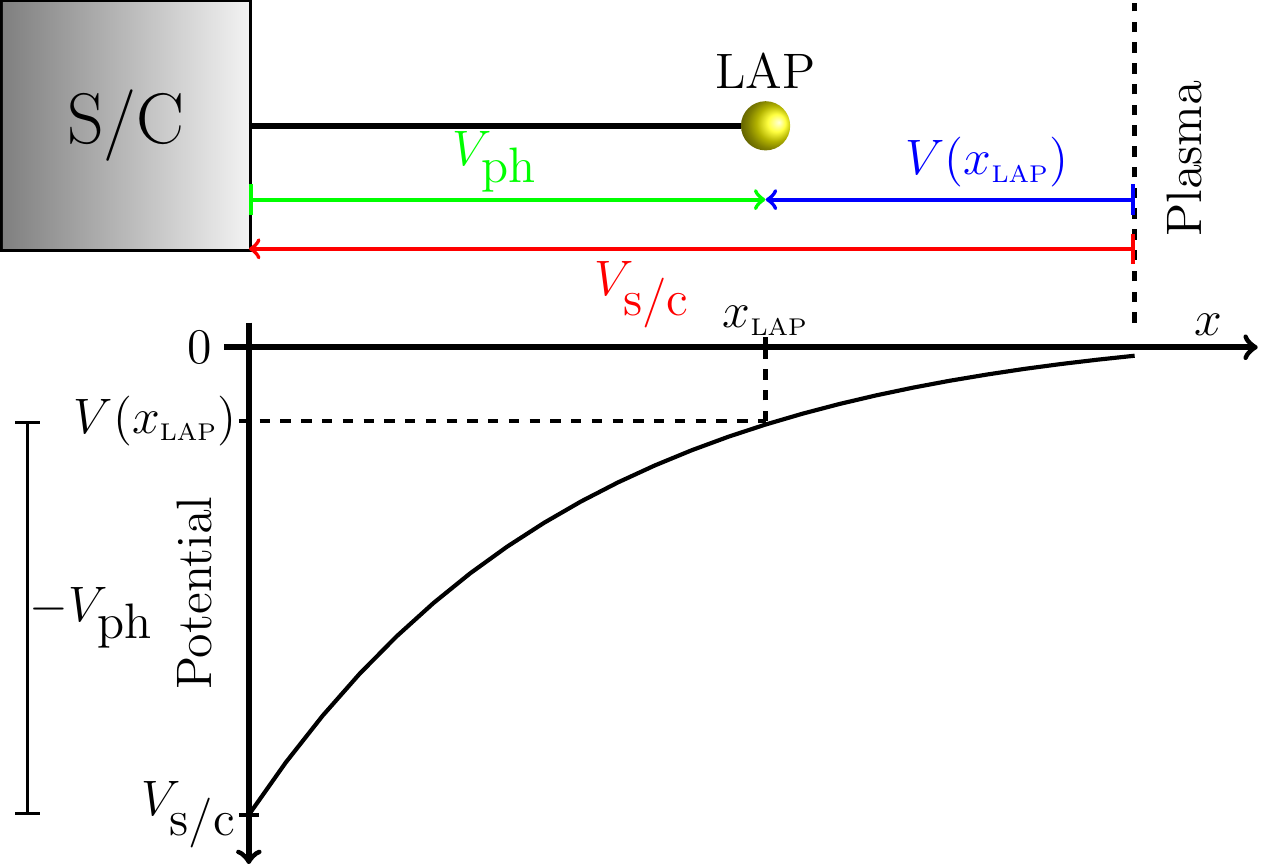}
\caption{Illustration of the potential structure around a (negatively) charged spacecraft. The local plasma potential observed by LAP at a distance $x_\textrm{LAP}$ from the spacecraft main body differs by an amount $V(x_{\textrm{LAP}})$ from the plasma potential at infinity due to the residual potential field of the spacecraft.}
\label{fig:potentials}
\end{figure}

LAP uses the spacecraft main body as ground, so the probe potential w.r.t.\ the ambient plasma at infinity is $V_\textrm{b} + V_{\textrm{s/c}}$. If the probe is located within a few Debye lengths of the spacecraft, the potential field of the charged spacecraft will persist at the location of the probe, as shown in Figure (\ref{fig:potentials}) where $V(x_\textrm{LAP})$ denotes the local plasma potential at the position of the probe, a distance $x_\textrm{LAP}$ away from the spacecraft main body. When the probe is biased to a negative potential w.r.t.\ the the local plasma potential, i.e.\ $V_\textrm{b} + V_{\textrm{s/c}} < V(x_\textrm{LAP})$, all the photoelectrons emitted from its surface are repelled away from the probe and contribute to the probe photoemission current, which hence saturates at a value independent of the actual value of the probe potential. When the probe becomes positive w.r.t.\ the surrounding plasma, i.e.\ $V_\textrm{b} + V_{\textrm{s/c}} > V(x_\textrm{LAP})$, a portion of the emitted photoelectrons are attracted back to the probe and the photoemission current decreases exponentially with increasing probe potential. This gives rise to a sharp inflection point in the sweeps, typically referred to as the photoelectron knee, at the bias potential $V_\textrm{b}$ for which $V_\textrm{b} + V_{\textrm{s/c}} = V(x_\textrm{LAP})$. Henceforth, this bias potential is denoted $V_{\textrm{ph}}$ and its position is annotated in Figure \ref{fig:IVcurve} (green circle). Thus we have

\begin{equation}
	V_{\textrm{s/c}} = V(x_\textrm{LAP}) - V_{\textrm{ph}} \quad,
\label{eq:Vsc_from_Vph}
\end{equation}
where $V_{\textrm{ph}}$ is the actually measured quantity and $V(x_\textrm{LAP})$ is generally unknown, but clearly depends on $V_{\textrm{s/c}}$. Assuming a linear relationship, $V(x_\textrm{LAP}) = (1 - \alpha) V_{\textrm{s/c}}$, gives

\begin{equation}
	V_{\textrm{s/c}} = (1-\alpha) V_{\textrm{s/c}} - V_{\textrm{ph}} \quad \Rightarrow \quad V_{\textrm{s/c}} = - \frac{V_{\textrm{ph}}}{\alpha} \quad,
\label{eq:Vph}
\end{equation}
i.e.\ this method only picks up some fraction $\alpha$ of the full spacecraft potential $V_{\textrm{s/c}}$.

The photoelectron knee in the bias sweep for a sunlit probe is detected in the analysis as a local maximum in the derivative and a peak in the second derivative through a central derivative scheme with a modest Savitzky-Golay algorithm noise filter in order not to distort the position of the knee. Errors in this technique 
arise from the filtering and the noise level of the LAP instrument, but also from plasma variations on comparable timescales as the sweep. 

\subsubsection{$V_{\textrm{s/c}}$ from a floating probe}
\label{sec:floating_probe}

$V_{\textrm{s/c}}$ measurements can also be obtained from a floating probe, i.e.\ a probe that is disconnected from the bias circuitry and therefore cannot carry a net current. In this case the probe will charge to the so called floating potential $V_\textrm{F}$, at which the currents naturally sum to zero. $V_\textrm{F}$ can be sampled at high time resolution (since no stepping of the bias potential is required) and multi-point measurements with floating probes allow for measuring electric fields\footnote{in sufficiently dense ($\gtrsim$10$^2$ cm$^{-3}$) plasmas, cf.\ \citet{Maynard1998a}.}, not possible with the bias potential sweeps described in the previous Section. $V_{\textrm{s/c}}$ can be obtained from such measurements in the following way: The spacecraft is floating ground, so the measured quantity is ${\Delta V = V_\textrm{F} - V_{\textrm{s/c}}}$, where $V_\textrm{F}$ is the floating potential of the probe w.r.t.\ the ambient plasma at infinity. The probe floating potential w.r.t.\ the local plasma potential at the position of the probe, $\delta V = V_\textrm{F} - V(x_\textrm{LAP})$, is typically a few volts in magnitude, corresponding to the typical temperature of ambient plasma electrons or emitted photoelectrons for positive and negative $\delta V$, respectively. Thus, $V_{\textrm{s/c}}$ can be obtained from $\Delta V$ as

\begin{multline}
	V_{\textrm{s/c}} = V_\textrm{F} - \Delta V = V(x_\textrm{LAP}) + \delta V - \Delta V = (1 - \alpha) V_{\textrm{s/c}} + \delta V - \Delta V \\
	\Rightarrow V_{\textrm{s/c}} = - \frac{\Delta V - \delta V}{\alpha} \quad .
\label{eq:VF}
\end{multline}
Since $\delta V$ is expected to be independent of $V_{\textrm{s/c}}$, the fraction of $V_{\textrm{s/c}}$ that is picked up by a floating probe is the same as the one picked up by bias potential sweeps, with the addition of an intercept $\delta V/\alpha$ on the order of a few volts (c.f.\ Section \ref{sec:evolution}).

While LAP bias potential sweeps can only be carried out once every 32 s at best, potential measurements from a floating probe can be obtained at a sampling frequency of up to 60 Hz, thus greatly increasing the time resolution of $V_{\textrm{s/c}}$ measurements. Although bias potential sweeps are much more prevalent throughout the mission, the latter kind of $V_{\textrm{s/c}}$ measurements are included here because their higher time resolution and unprocessed nature, being unaffected by any possible flaws or inaccuracies of the sweep analysis used to find $V_{\textrm{ph}}$, is advantageous for comparison to ICA data.

\subsubsection{Double-probe considerations}
\label{sec:double_probes}

For electric field measurements, both probes LAP1 and LAP2 float, producing simultaneous measurements of the floating potentials of both probes. Since the probes are mounted on booms of different lengths, they will capture different fractions of the spacecraft potential. However, in the absence of external electric fields, their floating potentials relative to the spacecraft and/or the surrounding plasma will in fact be virtually equal. To show this, note that the probe floating potential w.r.t.\ the local plasma potential at the position of the probe, $V_\textrm{F} - V(x_\textrm{LAP})$, is analogous to $V_{\textrm{s/c}}$ and can, for negative potentials $V_\textrm{F} - V(x_\textrm{LAP}) < 0$, be obtained from Equation (\ref{eq:Vsc}) with $A_\textrm{s/c}$ and $I_\textrm{ph}^\textrm{s/c}$ replaced by the total current collecting area of the probe $A_\textrm{p}$ and the probe photoemission current $I_\textrm{ph}^{p}$, respectively, while the electron density at the position of the probe $n_\textrm{e}^\textrm{p}$ is assumed to be reduced w.r.t.\ the ambient plasma due to the persisting potential field of the spacecraft by the Boltzmann factor, i.e.\

\begin{equation}
	n_\textrm{e}^\textrm{p} = n_\textrm{e} \exp \left\{ \frac{eV(x_\textrm{LAP})}{k_{\textrm{B}}T_\textrm{e}} \right\}.
\label{eq:Boltzmann}
\end{equation}
This gives

\begin{multline}
	V_\textrm{F} - V(x_\textrm{LAP}) = -\frac{k_{\textrm{B}} T_\textrm{e}}{e} \log \left\{ \frac{A_\textrm{p} n_\textrm{e}}{I_\textrm{ph}^\textrm{p}} \sqrt{\frac{k_{\textrm{B}} T_\textrm{e}}{2\pi m_\textrm{e}}} \exp \left\{ \frac{eV(x_\textrm{LAP})}{k_{\textrm{B}}T_\textrm{e}} \right\} \right\} \\ = -\frac{k_{\textrm{B}} T_\textrm{e}}{e} \log \left\{ \frac{A_\textrm{p} n_\textrm{e}}{I_\textrm{ph}^\textrm{p}} \sqrt{\frac{k_{\textrm{B}} T_\textrm{e}}{2\pi m_\textrm{e}}} \right\} - V(x_\textrm{LAP}).
\label{eq:probeVf}
\end{multline}
The first term on the right-hand-side of Equation (\ref{eq:probeVf}) is just the unperturbed floating potential $V_{\textrm{F},\infty}$ of a probe at infinity, beyond the influence of the spacecraft potential field. Thus we have

\begin{equation}
	V_\textrm{F}(x_\textrm{LAP}) = V_\textrm{F}(\infty) \quad \forall \quad x_\textrm{LAP} > 0.
\label{eq:mVf}
\end{equation}
Thus, the potential of the floating probe is driven positive due to the reduction of $n_\textrm{e}$ in the potential field of the spacecraft, but the presence of this field also lowers $V_\textrm{F}$ w.r.t.\ the ambient plasma at infinity by exactly the same amount.

When $V_\textrm{F} - V(x_\textrm{LAP}) > 0$, the situation is complicated by three new effects: i) the photoemission current $I_\textrm{ph}^\textrm{p}$ is no longer independent of the probe potential, since a portion of the emitted photoelectrons don't have enough energy to escape the attractive potential field of the probe and contribute to the net photoemission current. The relationship between $I_\textrm{ph}^\textrm{p}$ and the probe potential is then dependent on the probe size and geometry. ii) The electron current $I_\textrm{e}$ to the probe is no longer proportional to the repelling Boltzmann factor as in Equation (\ref{eq:Ie}). Also this depends on the probe geometry. iii) A potential barrier generally forms outside of the probe, as the strong inward potential gradient close to the positively charged probe is overtaken by the positive gradient in the potential field of the negatively charged spacecraft at larger distances.

Following the treatment of \citet{Olson2010}, we assume that the probe collects electrons from the top of the potential barrier, where the potential is $V_\textrm{bar}$ and the electron density is assumed to be reduced from the ambient value at infinity by the Boltzmann factor $\exp \left\{ eV_\textrm{bar}/k_{\textrm{B}}T_\textrm{e} \right\}$. Since $V_\textrm{F} > V_\textrm{bar}$, the OML formula for attractive potentials gives \citep{Olson2010}

\begin{equation}
	I_\textrm{e} = A_\textrm{p} n_\textrm{e} e \sqrt{\frac{k_{\textrm{B}} T_\textrm{e}}{2\pi m_\textrm{e}}} \exp \left\{ \frac{eV_\textrm{bar}}{k_{\textrm{B}} T_\textrm{e}}\right\} \left( 1 + \frac{e(V_\textrm{F} - V_\textrm{bar})}{k_{\textrm{B}} T_\textrm{e}} \right).
\label{eq:OML}
\end{equation}
If we assume that the photoemission current is made up of precisely those photoelectrons that have sufficient energy to overcome the potential barrier $V_\textrm{F} - V_\textrm{bar}$ and use the formula for photoemission current from a small spherical probe \citep{Grard1973}, we get

\begin{equation}
	I_\textrm{ph}^\textrm{p} = I_{\textrm{ph},0}^\textrm{p} \exp \left\{ -\frac{e(V_\textrm{F} - V_\textrm{bar})}{k_{\textrm{B}}T_\textrm{ph}} \right\} \left( 1 + \frac{e(V_\textrm{F} - V_\textrm{bar})}{k_{\textrm{B}}T_\textrm{ph}} \right),
\label{eq:Iph}
\end{equation}
where $I_{\textrm{ph},0}^\textrm{p}$ and $T_\textrm{ph}$ are the photosaturation current, equal to the photoemission current from a negative probe, and the temperature of the photoelectrons, respectively. Equating Equations (\ref{eq:OML}) and (\ref{eq:Iph}), assuming $T_\textrm{ph} = T_\textrm{e}$ and solving for $V_\textrm{F}$ yields

\begin{equation}
	V_\textrm{F} = -\frac{k_{\textrm{B}} T_\textrm{e}}{e} \log \left\{ \frac{A_\textrm{p} n_\textrm{e}}{I_{\textrm{ph},0}^\textrm{p}} \sqrt{\frac{k_{\textrm{B}} T_\textrm{e}}{2\pi m_\textrm{e}}} \right\}.
\end{equation}
Thus, again we find $V_\textrm{F} (x_\textrm{LAP}) = V_\textrm{F}(\infty)$, for all $x_\textrm{LAP} > 0$.

\subsubsection{Spacecraft ion current from Langmuir probe bias potential sweeps}
\label{sec:ion_current}

In addition to $V_{\textrm{ph}}$, the derivative of the attracted-ion current $dI_\textrm{i}/dV_\textrm{b}$ can usually also be reliably and consistently identified in the Langmuir probe sweeps \citep{Vigren2017}. The slope of the total sweep current at the lowest 10 V of a sweep, where only the ion current contributes to the slope, is typically used for this purpose. This ion slope then can be used to obtain a rough estimate of the ion current to the spacecraft. Applying the OML formulas for current collection of positive ions at negative potentials, for spherical and cylindrical geometry \citep{Allen1992}, respectively, for the probe and spacecraft gives

\begin{equation}
	I_\textrm{i,p} = -A_\textrm{p}n_\textrm{i}e\sqrt{\frac{k_{\textrm{B}}T_\textrm{i}}{2\pi m_\textrm{i}}} \left( 1 - \frac{eV_\textrm{b}}{k_{\textrm{B}}T_\textrm{i}} \right) \quad \textrm{(spherical probe)}
\label{eq:spherical_probe}
\end{equation}
	
\begin{equation}
	I_\textrm{i,S/C} = -A_{\textrm{S/C}} n_\textrm{i} e \sqrt{\frac{k_{\textrm{B}} T_\textrm{i}}{2\pi m_\textrm{i}}} \frac{2}{\sqrt{\pi}} \sqrt{1 - \frac{e V_{\textrm{S/C}}}{k_{\textrm{B}} T_\textrm{i}}} \quad \textrm{(cylindrical spacecraft)}
\label{eq:cylindrical_spacecraft}
\end{equation}
for singly charged ions of mass $m_\textrm{i}$, temperatue $T_\textrm{i}$, ion density $n_\textrm{i}$ at infinity and defining positive current away from the probe. Differentiating Equation (\ref{eq:spherical_probe}) w.r.t.\ $V_\textrm{b}$ and substituting into Equation (\ref{eq:cylindrical_spacecraft}) gives

\begin{equation}
	I_{i,\textrm{S/C}} = -\frac{2}{\sqrt{\pi}} \frac{A_{\textrm{S/C}}}{A_\textrm{p}} \frac{k_{\textrm{B}}T_\textrm{i}}{e} \frac{dI_{i,p}}{dV_\textrm{b}} \sqrt{1 - \frac{e V_{\textrm{S/C}}}{k_{\textrm{B}} T_\textrm{i}}}.
\label{eq:Isc}
\end{equation}

\subsubsection{Possible LAP2 surface contamination}

During parts of the mission, LAP2 behaved unexpectedly, including abnormal resistive, capacitive and emissive effects, most noticeably during the first months of the mission. This is likely due to contamination by a layer of condensed material, presumably of spacecraft origin, resulting from the long period ($\sim$2.5 years) of spacecraft hibernation before arrival at the comet, during which LAP2 was consistently in shadow. Its behaviour improved later on in the mission, but it is not clear too what extent abnormalities are still affecting the sweeps and the analysis for obtaining $V_\textrm{ph}$. For this reason, we only use spacecraft potential measurements from LAP1 in this Paper.

\subsection{Obtaining the spacecraft potential from RPC-ICA}
\label{sec:ICA}

RPC-ICA is a spherical top-hat electrostatic analyser (ESA) and magnetic momentum filter for ions \citep{Nilsson2007,Stenberg-Wieser2017}.
ICA is located on the spacecraft main body and oriented so that both the sun and nucleus directions are in its field of view during nominal spacecraft pointing. The instrument entrance is covered by a conductive grid grounded to the spacecraft main body. A deflection system behind the grid allows sweeping of the acceptance angle w.r.t.\ the instrument symmetry axis, the so called elevation angle, from 45$^\circ$ to 135$^\circ$. Accepted ions enter the ESA, consisting of two concentric hemispherical electrodes whose potentials are adjusted to accept only particles in a narrow energy range, a so called energy bin. The maximum energy range that the instrument can cover is from 5 eV/q to 40 keV/q and the energy resolution is $dE/E = 0.07$, except for energies below 30 eV/q where the effective energy resolution is reduced to $dE/E = 0.30$ because of pre-acceleration of particles into the ESA. For energies below 97 eV/q the inner electrode is held at a constant potential close to 0 V and the potential of the outer electrode is stepped between 0 and 10 V. For larger energies the outer electrode is held constant and the inner varied between 0 and 4 kV. After passing the ESA, ions are separated by mass in a circular magnetic field (the magnetic momentum filter) before hitting a micro-channel plate that registers ion impacts in each azimuthal and radial (i.e. mass) sector.

In the presence of detectable levels of locally produced ions, the spacecraft potential can be obtained from the lowest energy $E_{\textrm{\small ion,threshold}}$ of collected ions, since these have been accelerated by the spacecraft potential and gained an energy $-qV_{\textrm{s/c}}$, $q$ being the ion charge. Its location on the spacecraft main body means that unlike LAP, ICA captures the full value of $V_{\textrm{s/c}}$. However, the ICA energy spectra suffer from an unknown energy offset, due to a temperature-dependent offset in the high-voltage supply to the inner ESA electrode, typically on the order of 10 eV, but highly dependent on the instrument temperature \citep{Stenberg-Wieser2017}
. LAP $V_{\textrm{s/c}}$ measurements are needed for comparison to determine the value of this offset.


In this study, the energy threshold $E_{\textrm{th}}$ is identified in the ICA spectra as the lowest energy bin in which the number of counts is greater than or equal to five, and that is either followed by three energy bins with monotonically increasing number of counts, or for which any of the three following energy bins has a number of counts greater than or equal to 9. This has been found to produce a reasonable balance between sensitivity and robustness to noisy or otherwise perturbed spectra.

The nominal time resolution of ICA is 192 s. This is generally insufficient to track the spacecraft potential, which tends to vary rapidly in response to the highly variable and dynamic cometary plasma environment. By limiting the energy range to 5-97 eV/q and fixing the elevation angle close to 0$^\circ$, giving a field-of-view of 5x360$^\circ$ (a two dimensional measurement), ICA can be run in an operational mode with a 4-second time resolution that better captures the highly variable nature of the cometary plasma environment. This measurement mode is the only one of general practical use for cross-calibration of $V_{\textrm{s/c}}$ with LAP and only data from this particular mode is used in this Paper. The main measurement features and results produced by it is further discussed in a companion paper \citep{Stenberg-Wieser2017}.

The other ion instrument aboard Rosetta, the Ion and Electron Sensor (RPC-IES, \citet{Burch2007}), also an electrostatic analyser (for both ions and electrons) but without magnetic momentum filter, lacks a high-time-resolution mode of this kind and therefore does not provide as suitable data for comparison with LAP. Nevertheless, a comparison of ICA and IES data would certainly be of interest, but is deferred to future work.

\section{Results and discussion}

\subsection{LAP1 \& LAP2 comparison}

Double-probe floating potential measurements were carried out intermittently from May 2015 until EOM. Figure \ref{fig:Aug18} panel a) shows a 10-minute snapshot of measured probe floating potentials (w.r.t.\ the spacecraft, i.e.\ $\Delta V$, c.f.\ Section \ref{sec:Vsc_sweeps}) of LAP1 and LAP2 obtained on Aug 18, 2015, when Rosetta was in the northern hemisphere (Latitude $\approx 30^\circ$, Cheops system \citep{Preusker2015}) at a cometocetric distance of $\sim$340 km. The correspondence between the two probes is very good, as predicted by the theoretical analysis presented in Section \ref{sec:double_probes}. This is a ubiquitous feature of the floating potential measurements whenever the spacecraft potential is negative. For positive spacecraft potentials, e.g.\ encountered during the nightside excursion to $\sim$1500 km in late March to early April 2016 (c.f.\ Section \ref{sec:evolution} and Figure \ref{fig:lon_plot}) the correspondence is lost (not shown).


\subsection{LAP \& ICA comparison}

Observations by ICA in high time resolution (HR) mode are available intermittently from late April 2015 to late August 2016. Many of the measurements in April and  May 2015 suffered from an elevation binning problem \citep{Stenberg-Wieser2017}. In this study we therefore only consider measurements from the very end of May 2015 onwards, for which this problem did not occur.

The temperature dependence of the ICA energy offset has been determined to stop at about 13.5$^\circ$C \citep{Stenberg-Wieser2017}, above which the offset appears to be constant. In this study, we find that using this value as a constraint on sensor temperature is too restrictive since some of the most promising events for cross-calibration with LAP have slightly lower sensor temperatures and yet seem to be unaffected by temperature drift, possibly because the temperature is relatively constant during these events. We instead adopt a more forgiving constraint at 8.5$^\circ$C, which has been found to generally be more suitable for the bulk of the measurements used in this study.

We organise ICA HR data into blocks of consistent operational mode. This yields 172 data blocks of varying lengths, typically of a few hours. Data gaps in these blocks occasionally occur because of memory overflow in the internal buffer. 

\begin{figure*}
	\includegraphics[width=\textwidth]{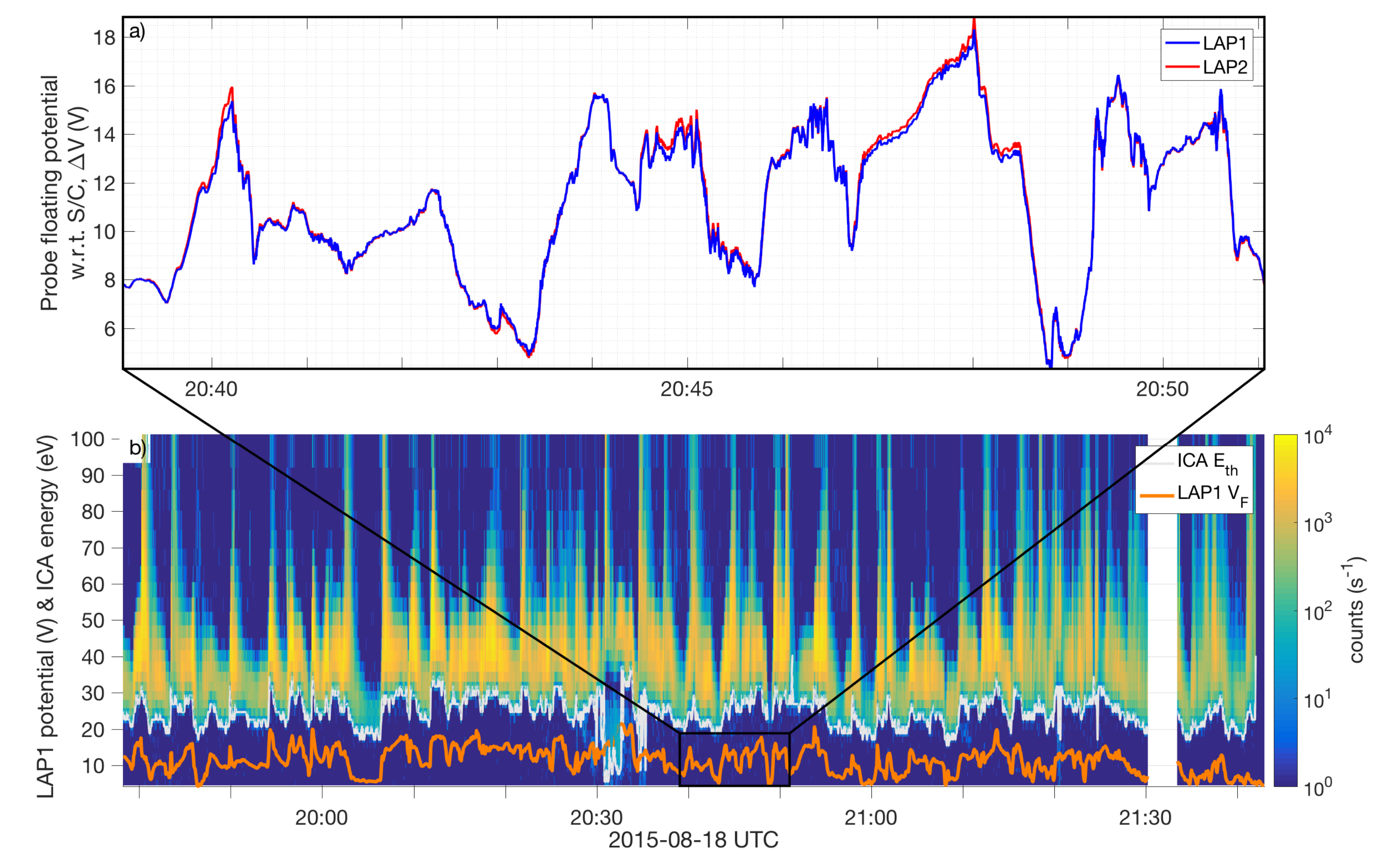}
\caption{\textbf{a)} Comparison of the floating potentials of LAP1 (blue) and LAP2 (red) w.r.t.\ the spacecraft ($\Delta V$, see text). The measured potentials are virtually identical, consistent with the simple model presented in Section \ref{sec:double_probes}, in the absence of significant electric fields in the plasma. \textbf{b)} ICA high time resolution ion energy spectra under stable sensor temperature of about 16$^\circ$C. LAP1 floating potential measurements (orange line) correlate well with the ion threshold energies (white line), validating their use as proxies for the spacecraft potential.}
\label{fig:Aug18}
\end{figure*}

Figure \ref{fig:Aug18} panel b) shows ICA ion spectra and LAP1 floating potential measurements during a period of high ICA time resolution and stable sensor temperature around 16$^\circ$C on August 18, 2015.
The LAP1 floating potential measurements (orange line in Figure \ref{fig:Aug18}), originally obtained at a sample frequency of 57.8 Hz, have been downsampled to the 4-second time-resolution of ICA by computing the average during each ICA sweep. The ICA energy threshold $E_{\textrm{th}}$, identified in the spectra as described in Section \ref{sec:ICA}, is also shown (white line). There is a very strong correlation between LAP1 $V_{\textrm{F}}$ and ICA $E_{\textrm{th}}$ in Figure \ref{fig:Aug18} (the Pearson correlation coefficient $R$, excluding the outliers, is about 0.9), confirming that they are in fact measuring the same thing and increasing our confidence in them as accurate estimators of the spacecraft potential. At about 20:30, some spurious counts in ICA below the spacecraft potential, the origin of which is still unknown, have been erroneously identified as $E_{\textrm{th}}$ by the algorithm. This illustrates that while generally accurate, the algorithm of Section \ref{sec:ICA} occasionally fails and tends to produce outliers in the data.

\begin{figure}
	\includegraphics[width=\columnwidth]{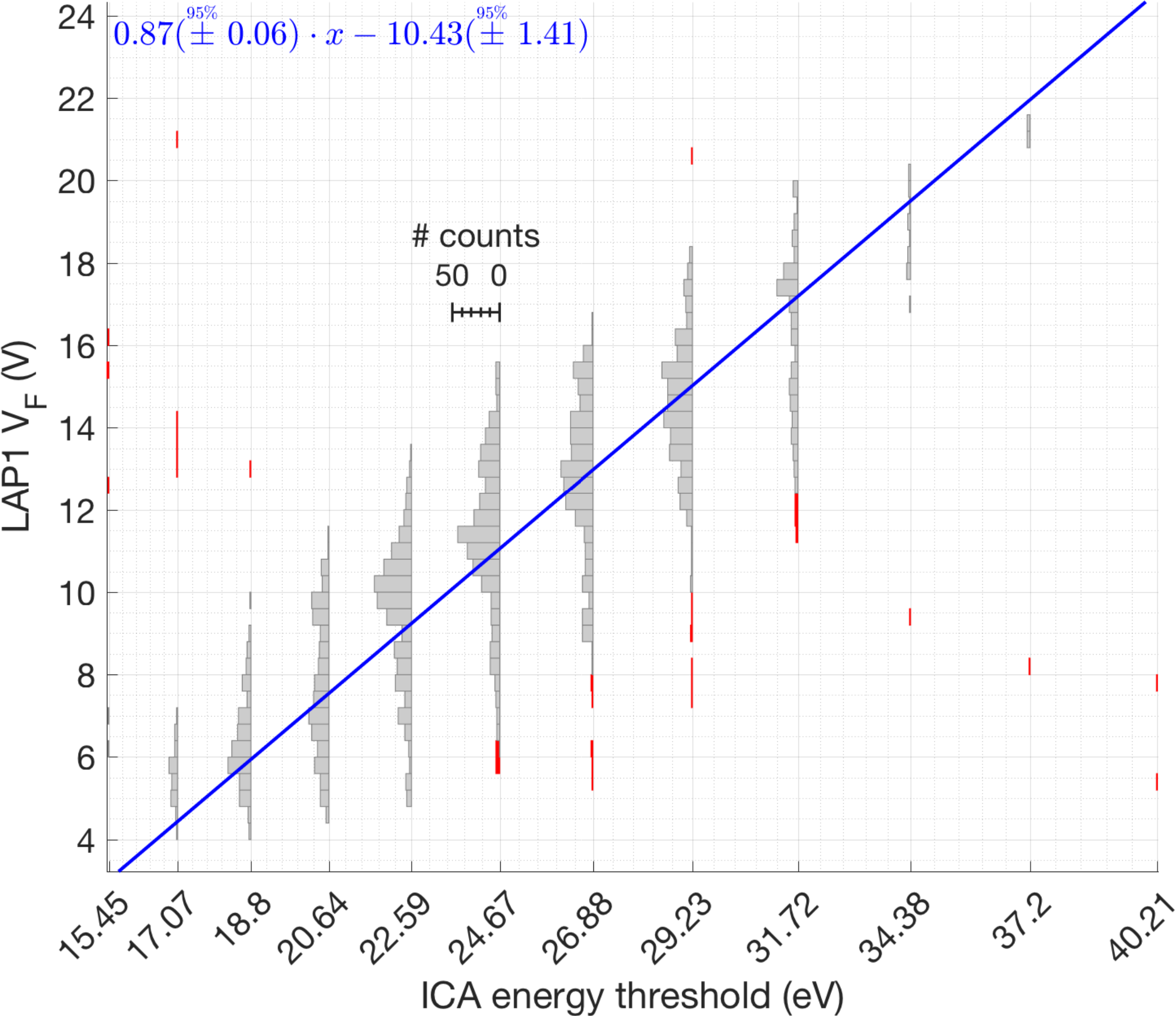}
\caption{LAP1 $V_{\textrm{F}}$ vs.\ ICA $E_{\textrm{th}}$ for the ICA HR block from August 18, 2015 shown in Figure \ref{fig:Aug18}. A linear relationship is evident between $V_{\textrm{F}}$ and $E_{\textrm{th}}$, the slope of which indicates that LAP1 $V_{\textrm{F}}$ picks up about 87\% of $V_{\textrm{s/c}}$.}
\label{fig:Aug18scatter}
\end{figure}

The relationship between $V_{\textrm{F}}$ and $E_{\textrm{th}}$ is illustrated graphically in Figure \ref{fig:Aug18scatter}. The large number of points in combination with the discrete nature of the ICA energy bins would make for a somewhat unintelligible figure if scatter-plotted in an ordinary fashion. Therefore, in order to clarify the distribution of measurement points in the plot, in Figure \ref{fig:Aug18scatter} the LAP1 $V_{\textrm{F}}$ measurements are shown as vertically binned histograms for each ICA energy bin. The left-ward extent of each histogram bar indicates the number of measurement points obtained in the $V_{\textrm{F}}$ interval corresponding to the vertical placement and extent of that bar and in the corresponding ICA energy bin. 
A linear relationship between $V_{\textrm{F}}$ and $E_{\textrm{th}}$ is evident in Figure \ref{fig:Aug18scatter}. A total least squares (TLS) linear fit \citep{vanHuffel1991} of $V_{\textrm{F}}$ to $E_{\textrm{th}}$ is shown (blue line), the equation for which is shown in the upper left corner. Also shown are 95\% confidence intervals based on the estimator covariances given by Equation (8.47) in \citet{vanHuffel1991}. When computing the TLS fit, which as opposed to the ordinary least squares (OLS) fit also takes into account errors in the independent variable but is more sensitive to outliers, a generalised extreme Studentised deviate (ESD) test \citep{Rosner1983} is iteratively performed on sequential TLS fits until no outliers (w.r.t.\ a normal distribution and at a significance level of 0.01) are found in the fit residuals. Outliers found by this method, and hence excluded from the final TLS fit, are coloured red in Figure \ref{fig:Aug18scatter}. Some 40-50 additional outliers at ICA energies between about 5-15 eV and $V_{\textrm{F}}$ in the range 12-22 V, originating from the spurious ICA counts below $V_{\textrm{s/c}}$ in Figure \ref{fig:Aug18}), have also been identified by the method, but have been cropped out of the figure for improved readability. This procedure has been found to generally produce accurate linear fits that are robust to outliers and yield reasonable confidence intervals for the bulk of the $V_{\textrm{F}}$-to-$E_{\textrm{th}}$ comparisons performed in this study. The slope of the regression line gives the fraction of $V_{\textrm{s/c}}$ that is picked up by LAP1 floating potential measurements, in this case found to be about 0.9. The intercept is a convolution of the ICA energy offset and an offset due to the floating potential of LAP1 w.r.t.\ its local plasma potential, $\delta V/\alpha$, discussed in Section \ref{sec:double_probes}, and should thus not be confused for the pure ICA offset.

\begin{figure*}
	\includegraphics[width=\textwidth]{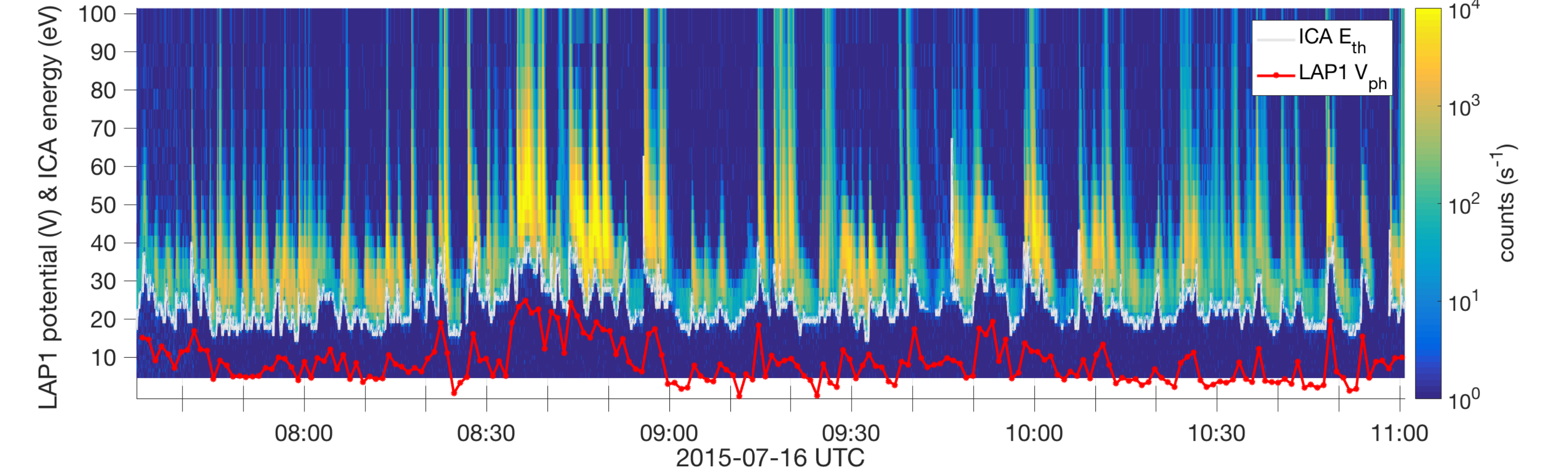}
\caption{ICA high time resolution ion energy spectra under sensor temperatures above 13$^\circ$C. LAP1 spacecraft potential measurements (red line) correlate well with the ion threshold energies (white line), validating their use as proxies for the spacecraft potential.}
\label{fig:Jul16}
\end{figure*}

Figure \ref{fig:Jul16} shows another example from July 16, 2015, when LAP1 was performing sweeps with a cadence of 64 seconds. Each sweep takes less than one second so the $V_{\textrm{ph}}$ measurements (red line) should be interpreted as sparse snapshots rather than smoothing time-averages of the spacecraft potential. As before, the white line is the ICA energy threshold $E_{\textrm{th}}$ as identified by the algorithm described in Section \ref{sec:ICA}. In spite of the sparsity of $V_{\textrm{ph}}$ measurements, the correlation between $V_{\textrm{ph}}$ and $E_{\textrm{th}}$ is very good during this time interval ($R \approx 0.9$ in this case as well).

\begin{figure}
	\includegraphics[width=\columnwidth]{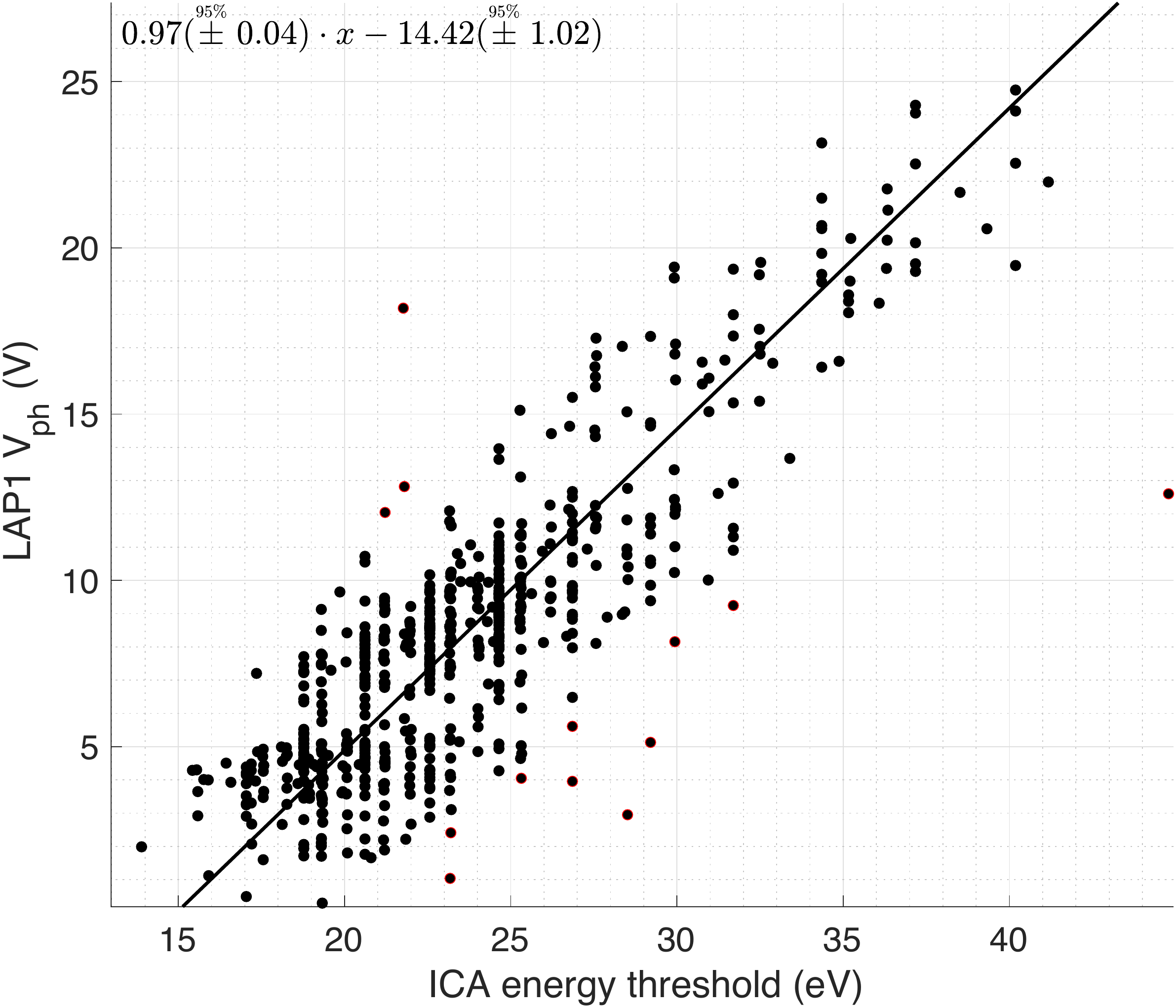}
\caption{LAP1 $V_{\textrm{ph}}$ vs.\ ICA $E_{\textrm{th}}$ for the ICA HR block from July 16, 2015 shown in Figure \ref{fig:Jul16}. A linear relationship is evident between $V_{\textrm{ph}}$ and $E_{\textrm{th}}$, the slope of which indicates that LAP1 $V_{\textrm{ph}}$ picks up about 90\% of $V_{\textrm{s/c}}$.}
\label{fig:Jul16scatter}
\end{figure}

The data shown in Figure \ref{fig:Jul16} comes from a rather long ICA HR data block, over 17 hours, with LAP1 sweeps throughout. Only about 3.5 hours from this block are shown in Figure \ref{fig:Jul16} in order for the short-timescale variations to be clearly discernible. Figure \ref{fig:Jul16scatter} shows a scatter plot of $V_{\textrm{ph}}$ vs.\ $E_{\textrm{th}}$ for the entire data block, where the $E_{\textrm{th}}$ measurements have been linearly interpolated to the $V_{\textrm{ph}}$ sample times (LAP1 mid-sweep times).
A clearly linear relationship between $V_{\textrm{ph}}$ and $E_{\textrm{th}}$ can be observed, quantified by a linear TLS fit, shown as the solid black line in Figure \ref{fig:Jul16scatter} and the equation for which is shown in the upper left corner. The same procedure described above is used to detect and exclude outliers; these are marked by red edges in Figure \ref{fig:Jul16scatter}. Most often, they originate from erroneous identification of $V_{\textrm{ph}}$ in the LAP sweeps (to be discussed further below) although errors in the identification of $E_{\textrm{th}}$ in the ICA spectra sometimes also contribute.

The estimator covariances used for calculations of confidence intervals when LAP1 was performing floating potential measurements are generally less accurate for the reduced number of samples available when LAP1 is in sweep mode. Therefore, confidence intervals are instead computed using bootstrapping \citep{Efron1994} with 1000 bootstrap samples. Such confidence intervals for the slope and intercept are shown along with the regression line equation in Figure \ref{fig:Jul16scatter}. The slope of the regression line gives the fraction of $V_{\textrm{s/c}}$ that is picked up by LAP1 $V_{\textrm{ph}}$ measurements, in this case found to be close to unity. The $V_{\textrm{ph}}$ measurements are expected to be bereft of any offset and the intercept can thus be identified as the ICA energy offset (with reversed sign), in this case thus found to be about 14 eV.

\begin{figure}
	\includegraphics[width=\columnwidth]{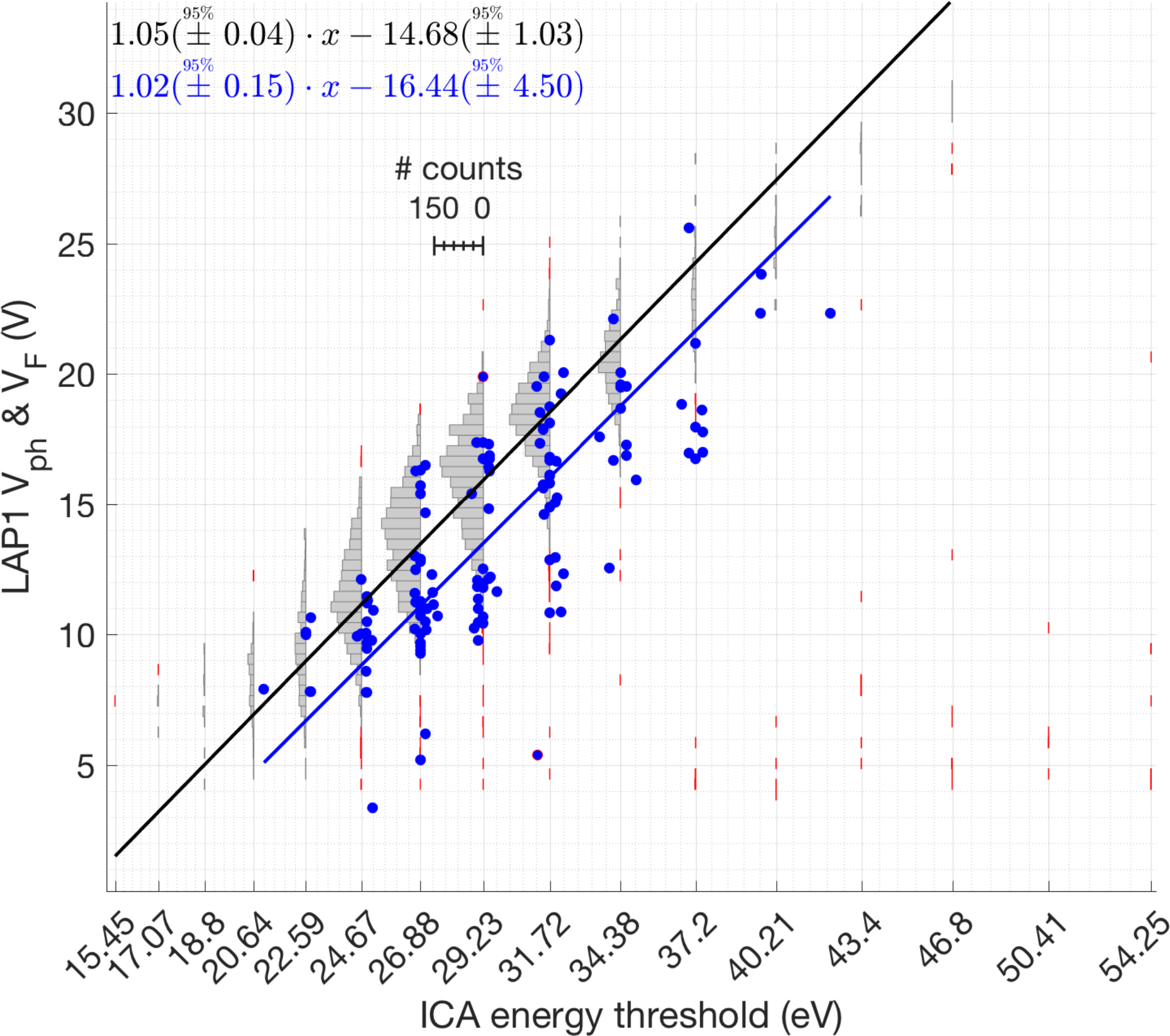}
\caption{LAP1 $V_{\textrm{F}}$ (gray/red histogram bars) and $V_{\textrm{ph}}$ (blue scatter points) vs.\ ICA $E_{\textrm{th}}$ during an ICA HR block on January 20-21, 2016. Similar slopes of the TLS regression lines (black and blue, respectively) indicate that $V_{\textrm{F}}$ and $V_{\textrm{ph}}$ pick up the same fraction of $V_{\textrm{s/c}}$, as predicted theoretically in Section \ref{sec:floating_probe}.}
\label{fig:Jan20scatter}
\end{figure}

Figure \ref{fig:Jan20scatter} shows similar scatter plots as in Figures \ref{fig:Aug18scatter} and \ref{fig:Jul16scatter} for an ICA HR data block between about 22:00 on January 20 and 15:00 on January 21, 2016. During the first half of this block LAP1 performed floating potential measurements and during the second half sweeps, allowing for a direct comparison of the two methods under similar conditions. TLS regression lines have been computed for each type of measurement using the respective procedures described above; these are shown in Figure \ref{fig:Jan20scatter} as black ($V_{\textrm{F}}$) and blue ($V_{\textrm{ph}}$) lines, with their equations (correspondingly coloured) shown in the upper left corner. Clearly, the slopes are the same, meaning that both methods pick up the same fraction of $V_{\textrm{s/c}}$, confirming the theoretical predictions in Section \ref{sec:floating_probe}. The intercepts are not quite the same, on account of the offset present in the $V_{\textrm{F}}$ measurements discussed above. The $V_{\textrm{ph}}$ measurements, presumed to be offset-free as previously discussed, suggest an ICA energy offset close to 16 eV for this data block, i.e.\ somewhat higher than previously found (Figure \ref{fig:Jul16scatter}) but the difference is not statistically significant.

For the $V_{\textrm{ph}}$ TLS regression in Figure \ref{fig:Jan20scatter}, the confidence intervals are quite large, owing to the the large scatter in the data. This is quite typical of the $V_{\textrm{ph}}$ measurements in this study and is likely due to a combination of the sparsity of these measurements and errors in the algorithm used to identify the photoelectron knee in the sweeps (c.f.\ Section \ref{sec:Vsc_sweeps}). This algorithm, which is used throughout the entire LAP data set, is quite complex since it has to deal with a large number of different sweeps with different characteristics, noise and disturbances, and it is not always very accurate.

\begin{figure*}
	\includegraphics[width=\textwidth]{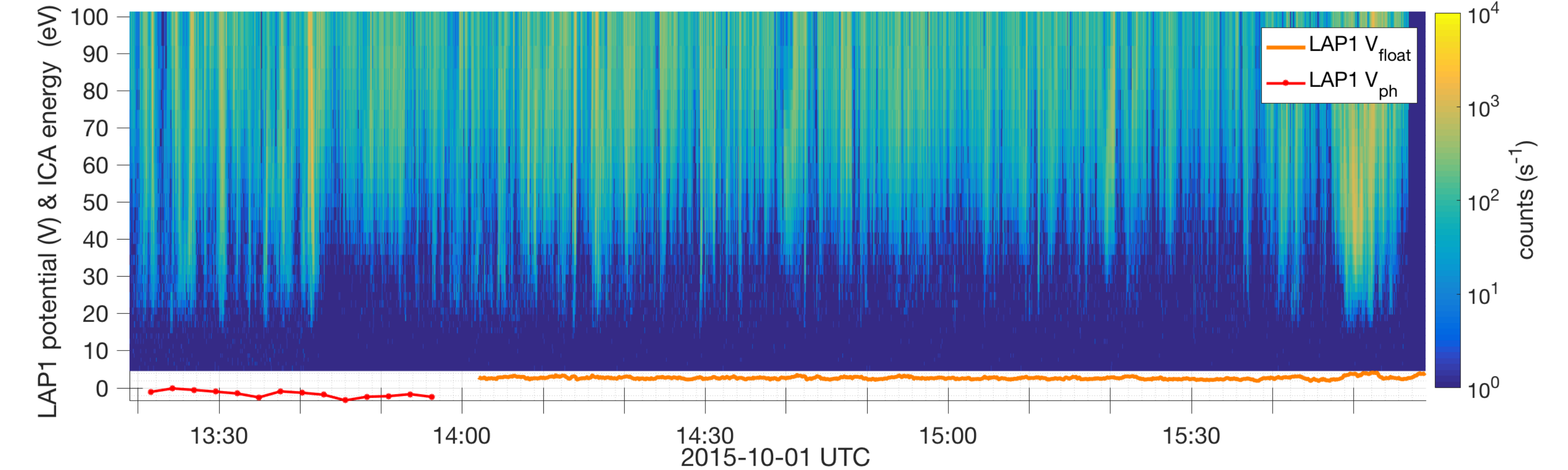}
\caption{ICA high time resolution ion energy spectra from October 1, 2015. LAP1 $V_{\textrm{ph}}$ (red line) and floating potential measurements (orange line) correlate well with the ion threshold energies in the beginning, when $V_{\textrm{F}}$ is quite large and $V_{\textrm{s/c}}$ thus clearly negative, but when $V_{\textrm{F}}$ becomes low, indicating a $V_{\textrm{s/c}} \gtrsim 0$~V, the correlation is lost.}
\label{fig:Oct1}
\end{figure*}

Figure \ref{fig:Oct1} shows an example from October 1, 2015, when LAP1 was initially performing sweeps but subsequently changed to floating potential mode. Here, the correlation between $V_{\textrm{F}}$ and $E_{\textrm{\small ion,threshold}}$ is lost altogether. The low-energy cutoff is much less sharp in this case and the algorithm for finding $E_{\textrm{\small ion,threshold}}$ generally produces very erratic results, thus these are not shown in this plot. It can be observed that the energies at which significant count rates start to appear are generally higher here than in the previously shown cases, in spite of the spacecraft potential being much weaker. In fact, from the (presumably offset-free) $V_{\textrm{ph}}$ measurements in the beginning of the shown interval, we can surmise that $V_{\textrm{s/c}}$ is close to zero and even slightly positive. We attribute the observed behaviour to a much weakened flux of locally produced ions, at least in the ICA field of view (which may depend on the spacecraft potential), probably due to them no longer being effectively accelerated into the instrument by a strong negative spacecraft potential, or possibly due local ionisation being lost altogether. Instead, the flux into the instrument is presumably dominated by an ion population produced further away from the spacecraft and then accelerated towards it by an ambient electric field. This kind of behaviour was most often encountered during the two spacecraft excursions in October 2015 and April 2016, when the distance to the nucleus was the largest, but also in May-June 2015 when the combination of relatively large cometocentric distance, low comet activity and seasonal effects tended to give a weakly and/or positively charged spacecraft (the behaviour of $V_{\textrm{s/c}}$ during various parts of the mission is discussed further in Section \ref{sec:evolution}). This is also when any loss of local production would be most likely to occur.

\begin{figure*}
	\includegraphics[width=\textwidth]{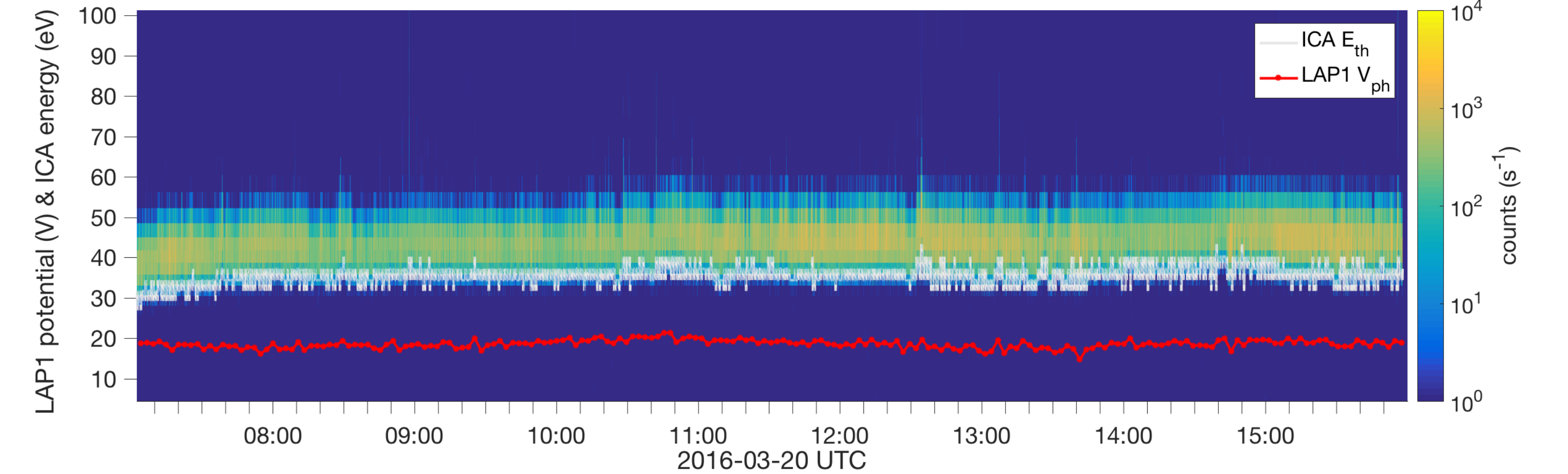}
\caption{ICA high time resolution ion energy spectra under sensor temperatures above 15$^\circ$C. The correlation between LAP1 spacecraft potential measurements (red line) and the ion threshold energies (white line) is weak during this time period, attributed to the poor dynamic range of $V_{\textrm{s/c}}$ and consequently low signal-to-noise ratio.}
\label{fig:Mar20}
\end{figure*}

Figure \ref{fig:Mar20} shows another example from March 20, 2016, this time when LAP1 was performing sweeps. This period was quite calm, with very small variations in $V_{\textrm{s/c}}$. Consequently, the signal variance was dominated by the measurement noise (including errors in the respective algorithms for identifying $V_{\textrm{ph}}$ and $E_{\textrm{th}}$) and the correlation was poor throughout ($R \approx 0.4$). This highlights the importance of sufficient dynamic range and good signal-to-noise ratios in order for the dual-instrument comparison to work.

\begin{figure}
	\includegraphics[width=\columnwidth]{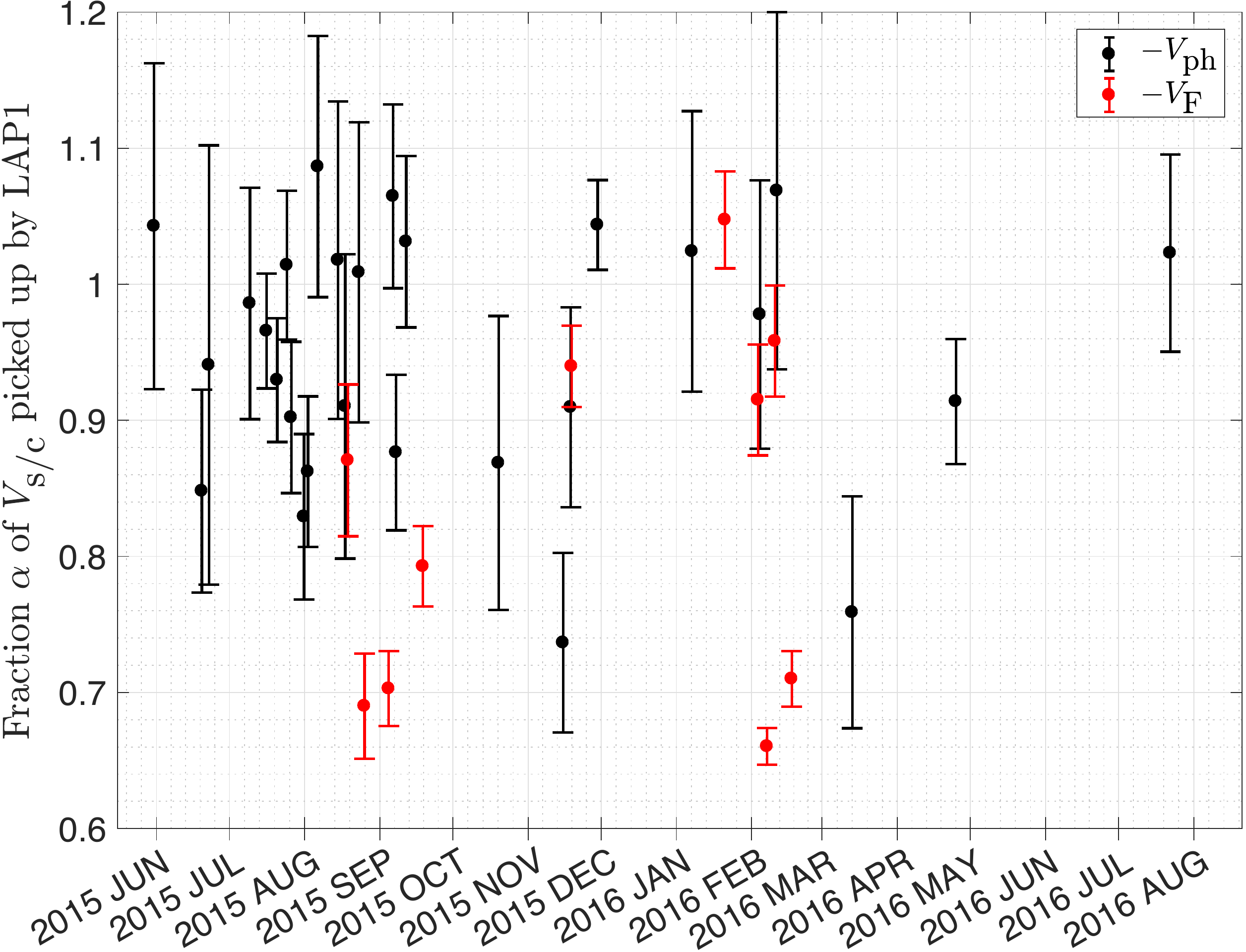}
\caption{Overview of the estimated fraction of $V_{\textrm{s/c}}$ picked up by LAP1 for all data blocks fulfilling the selection criteria (see text). Black and red points are $V_{\textrm{ph}}$ and $V_{\textrm{F}}$ measurements, respectively, and the errorbars show 95\% confidence intervals.}
\label{fig:overview}
\end{figure}

Finally, having examined a number of representative cases in Figures \ref{fig:Aug18}-\ref{fig:Mar20}, we expand the analysis to the entire available data set with the aim of better constraining the fraction $\alpha$ of $V_{\textrm{s/c}}$ picked up by LAP1 throughout the mission, as well as the ICA energy offset. The requirements of sufficiently high sensor temperature, strong fluxes of locally produced ions and good signal-to-noise ratios severely reduce the number of usable ICA HR data blocks. Identification of data blocks where cross-calibration of the instruments might be feasible is aided by the results of \citet{Stenberg-Wieser2017}, who found that the bulk of the ICA HR spectra collected throughout the mission could be grouped into 5 different types, which was done manually for every hour of collected data. We find that their Type 3 generally corresponds quite well to cases that we attribute to a weakened flux of locally produced ions and that are therefore unsuitable for cross-calibration. We therefore only include data blocks with more than one hour of data of type 1, 2, 4 or 5 at acceptable ($>8.5^\circ$) sensor temperature and for which the resulting number of cross-calibration points is at least 50 (this is not a problem for $V_{\textrm{F}}$ measurements, but sometimes for the much sparser $V_{\textrm{ph}}$). Finally, we also require that the correlation coefficient (excluding outliers) be at least 0.8 and that the number of outliers not be more that 15\% of the total number of regression points (also not an issue for $V_{\textrm{F}}$, but sometimes for $V_{\textrm{ph}}$). Figure \ref{fig:overview} shows an overview of the estimated fraction of $V_{\textrm{s/c}}$ picked up by LAP1 for all data blocks fulfilling these criteria. Black and red points are $V_{\textrm{ph}}$ and $V_{\textrm{F}}$ measurements, respectively, and the errorbars show 95\% confidence intervals.

The fraction of $V_{\textrm{s/c}}$ picked up by LAP1 is found to generally be between about 0.7 and 1, but highly variable in this range. For the $V_{\textrm{ph}}$ measurements the uncertainty is generally large. We attribute this to their sparsity and possible inaccuracies in their derivation from the sweeps. There is no obvious trend or correlation with heliocentric distance or position of the spacecraft w.r.t.\ the nucleus that could be used for straight-forward calibration of the entire $V_{\textrm{s/c}}$ data set. Presumably, the fraction picked up by LAP depends on local plasma parameters such as electron density and temperature, which determines the screening distance of electrostatic potentials in the plasma. These vary a lot on small temporal and spatial scales and are not generally determined with any accuracy in an automated fashion over longer time periods. A few points in Figure \ref{fig:overview} appear to be above unity with statistical significance, though just barely. We do not consider it plausible that LAP would pick up more than the full spacecraft potential, it seems more likely that instrumental effects in ICA might prohibit it from capturing the full dynamic range of $V_{\textrm{s/c}}$, or that the statistical analysis used here fails to account for the full uncertainty of the measurements.

\begin{figure}
	\includegraphics[width=\columnwidth]{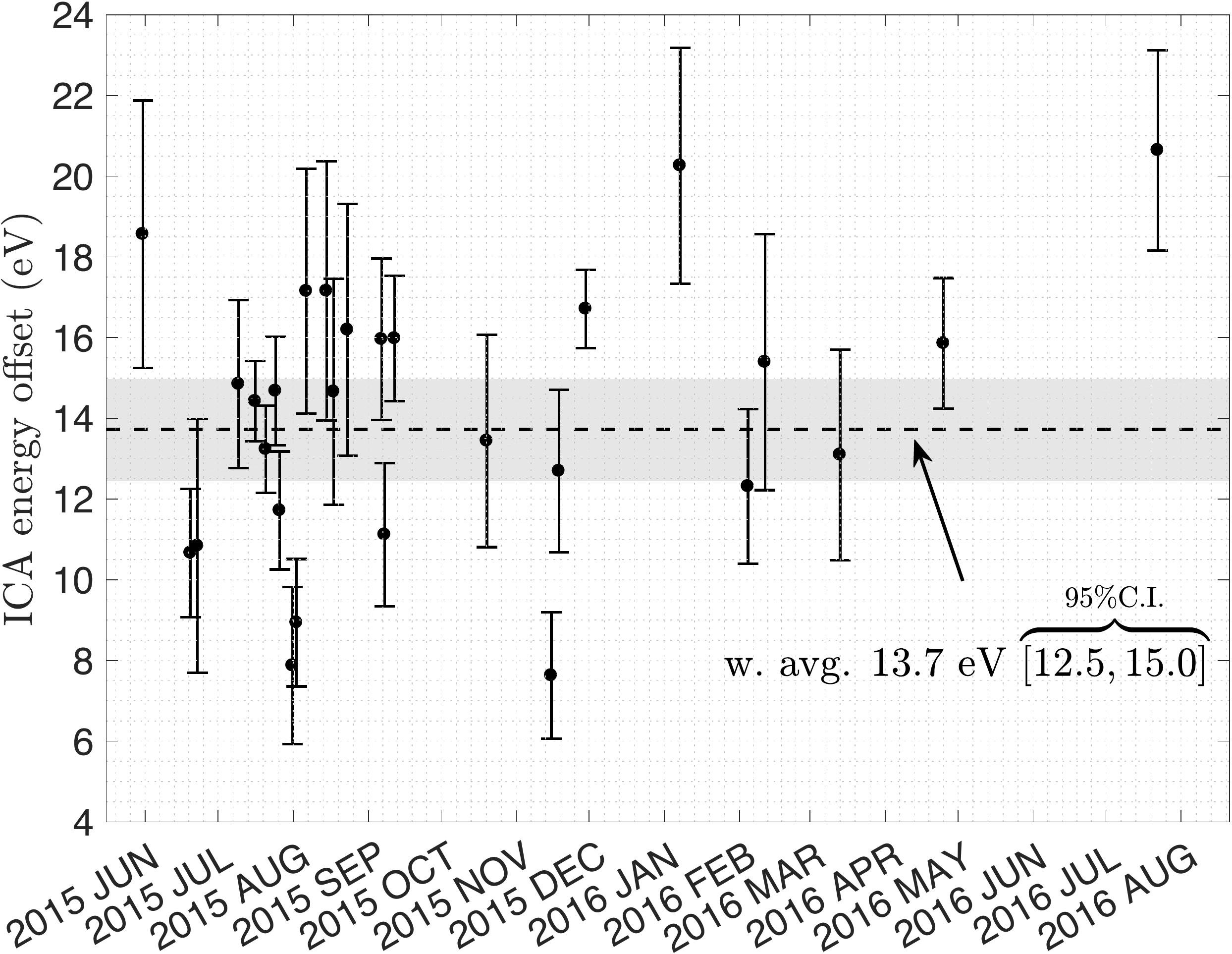}
\caption{overview of the estimated ICA energy offset for all data blocks of $V_{\textrm{ph}}$ measurements fulfilling the selection criteria (see text). Errorbars show 95\% confidence intervals. A weighted arithmetic mean of all values of 13.7 eV is also shown (dashed line), along with its 95\% confidence interval [12.5, 15.0] eV (shaded area).}
\label{fig:ICA_offset_overview}
\end{figure}

Figure \ref{fig:ICA_offset_overview} shows an overview of the estimated ICA energy offset for all data blocks of $V_{\textrm{ph}}$ measurements fulfilling the above criteria. As before, the errorbars show 95\% bootstrapped confidence intervals. The scatter is quite large, with values in the range 7 - 21 eV. Unlike the fraction $\alpha$ of $V_{\textrm{s/c}}$ picked up by LAP, which can be expected to vary substantially in response to changes in the ambient plasma environment, the ICA energy offset is expected to be constant throughout the mission (for the acceptable sensor temperatures used here). Therefore, we compute a weighted arithmetic mean of all values in Figure \ref{fig:ICA_offset_overview}, with each weight proportional to the reciprocal of the corresponding bootstrapped variance from the statistical analysis described above. The result, shown as a dashed horizontal line in Figure \ref{fig:ICA_offset_overview}, is 13.7 eV, with a bootstrapped confidence interval at the 95\% level (shaded area in Figure \ref{fig:ICA_offset_overview}) between 12.5 and 15.0 eV. It can be observed that close to a third of the samples (8 out of 27) have confidence intervals that do not overlap the confidence interval of the weighted mean and are thus statistically significantly different from it. This casts some doubt on the assumption that the energy offset is indeed constant, or that it is accurately obtained throughout the analysis used here. Another possibility is, again, that the statistical analysis used here fails to account for the full uncertainty of the measurements. Nevertheless, the 13.7 eV presented here represents the best estimate of the ICA energy offset to date and is used by e.g.\ \citet{Stenberg-Wieser2017} to refine the energy table and their ion temperature calculations.

\subsection{Ion current to the spacecraft}
\label{sec:Ii_SC}

\begin{figure*}
	\includegraphics[width=\textwidth]{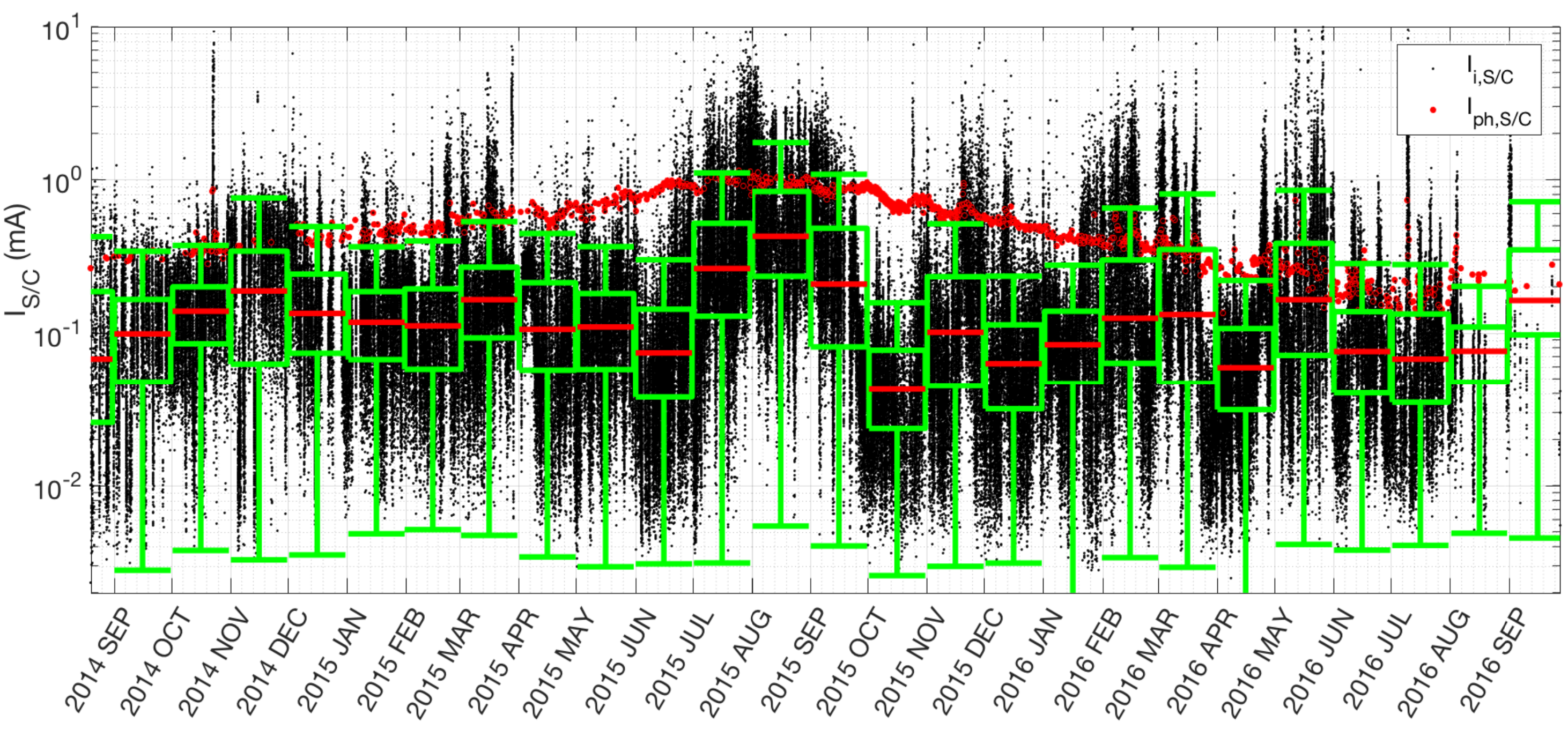}
\caption{Spacecraft ion current estimated from LAP1 sweep ion slope (black points) and photoemission current estimated from LAP1 photoemission currents (red dots). Box plots show monthly medians and 75\% and 99\% quantiles of the spacecraft ion current.}
\label{fig:I_iSC}
\end{figure*}

Figure \ref{fig:I_iSC} shows the ion current to the spacecraft calculated from the ion slope $dI_\textrm{i}/dV_\textrm{b}$ in LAP1 Langmuir probe bias voltage sweeps according to Equation (\ref{eq:Isc}), from late august 2014 to EOM (black dots). An ion temperature of 5 eV has been assumed throughout, in line with the most recent estimates of \citet{Stenberg-Wieser2017}. There is a lot of scatter due to the highly variable and dynamic nature of the cometary plasma. For this reason, monthly box plots are provided, for which the (red) line in the middle of each (green) box is the sample median and the tops and bottoms are the 25th and 75th percentiles of the samples, respectively. Whiskers are drawn to the furthest observations within 1.5 times the interquartile range of the top or bottom of each box, corresponding to approximately $\pm 2.7\sigma$ and 99.3\% coverage if the data are normally distributed \citep{McGill1978}. Also shown in Figure \ref{fig:I_iSC} is the estimated photoemission current to the spacecraft (red dots), calculated by rescaling the LAP1 photoemission current of \citet{Johansson2017} to the total illuminated area of the spacecraft, estimated here to be 70 m$^2$.

It is clear from Figure \ref{fig:I_iSC} that the spacecraft ion current was entirely negligible compared to the much larger photoemissison current for more than or close to 99\% of the measurements during most of the mission. Exceptions to this are primarily March and May 2016, when the close proximity to the nucleus gave somewhat stronger ion currents and there may have been a significant contribution of $I_\textrm{i,S/C}$ to $V_{\textrm{s/c}}$ in upwards of 25\% of the measurements.

The estimate of the spacecraft ion current is quite sensitive to the ion temperature. The above analysis holds quite well also up to $T_\textrm{i} = 10$ eV (not shown), but if the ion temperature becomes higher than that, $I_\textrm{i,S/C}$ becomes non-negligible compared to $I_{\textrm{ph,S/C}}$ for large parts of the mission.

\subsection{Evolution of the spacecraft potential}
\label{sec:evolution}

In this Section, we present an overview of the evolution of the spacecraft potential during virtually the entire stay of Rosetta at the comet, from early September 2014, when LAP1 was first sunlit and performing Langmuir probe sweeps on a regular basis, until EOM on September 30 2016. This is essentially a continuation of the analysis of the long term evolution of the spacecraft potential and cometary plasma environment of \citet{Odelstad2015}.

\begin{landscape}
	\begin{figure}
		\includegraphics[width=0.9\paperheight]{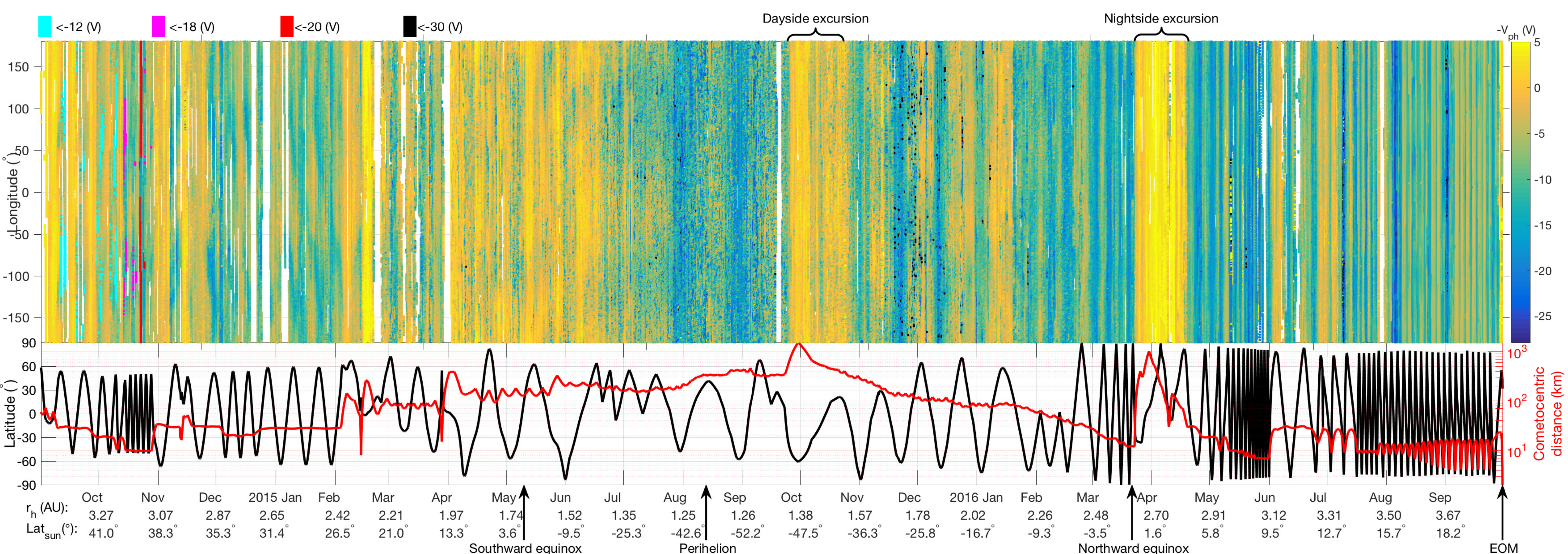}
	\caption{Overview of RPC-LAP spacecraft potential measurements from September 1, 2014 to the end of mission (EOM) on September 31, 2016. (top) The negative of the photoelectron knee potential color coded on a time-longitude map. Cyan, magenta, red and black colours indicate measurements where the negative of the photoelectron knee potential went below the measurement range of -12~V, -18~V, -20~V or -30~V, respectively. (bottom) The latitude in black (to be read off the left-hand vertical axis) and the cometocentric distance in red (right-hand vertical axis). Heliocentric distance and latitude of the sun in the comet-fixed frame \citep{Preusker2015} are shown below the timeaxis. Perihelion, southward and northward equinox and the two excursions are annotated for clarity.}
	\label{fig:lon_plot}
	\end{figure}
\end{landscape}

Figure \ref{fig:lon_plot} shows an overview of $V_\textrm{s/c}$ as gauged by the photoelectron knee potential (i.e.\ without any applied correction factor), plotted versus time on the horizontal axis, longitude on the vertical axis and with each point colour-coded by $-V_\textrm{ph}$ from LAP1 sweeps. The bottom panel shows the latitude in black and the cometocentric distance in red (to be read off the left-hand and right-hand vertical axes, respectively). The voltage range swept by the Langmuir probe varied between different measurement modes, the upper limit being either 12~V, 18~V, 20~V or 30~V. Occasionally, $V_\textrm{ph}$ would become larger than the upper edge of the sweep bias range; for these cases this upper edge is used as a lower limit. In Figure \ref{fig:lon_plot}, such limit values are colour-coded by cyan, magenta, red or black for maximum sweep potentials of 12~V, 18~V, 20~V and 30~V, respectively.

The period from the beginning of September 2014 to the end of March 2015 was examined by \citet{Odelstad2015}. During this period at relatively large heliocentric distance (3.5 - 2.1~AU), Rosetta was able to go very close to the nucleus ($\lesssim$ 30 km), a possibility that did not come again until May 2016. $V_\textrm{s/c}$ was negative within $\sim$50 km of the nucleus throughout this period, with a strong radial dependence and the the most negative spacecraft potentials ($-V_\textrm{s/c} \gtrsim 15 - 20$~V) observed in the northern (summer) hemisphere above the neck region of the comet nucleus. In this regard, the plasma density was found to trace the neutral gas density.

From early April to mid-June 2015 (2 - 1.5~AU) the spacecraft potential was generally in the interval between -5~V and + 5~V, only occasionally dipping down to about -10~V. During this time period, the solar latitude was close to $0^\circ$, with southward equinox on May 10 marking the end of northern summer and the beginning of southern summer. In April, the dependence of $V_\textrm{s/c}$ on spacecraft latitude is somewhat unclear, with lower spacecraft potentials, interpreted as increased plasma density, generally observed at both high and low latitudes and the most positive potentials observed near the comet equatorial plane. After equinox on May 10, it becomes clearer that the most negative potentials are observed above the southern hemisphere and the most positive above the northern hemisphere. However, the spacecraft potential still appears to go significantly more positive near the equatorial plane. This can be compared to the density of neutral water vapour, which \citet{Hansen2016} found to often have a minimum in the equatorial plane in this period, driven by active areas in the north and south.

Towards the end of June, as the heliocentric distance decreased below about 1.4~AU, a clear trend towards more negative spacecraft potentials began, with $V_\textrm{s/c}$ being mostly negative, intermittently dipping down to around -20~V. There was also a distinct longitude-modulation of the signal when the spacecraft was in the northern hemisphere, with more negative spacecraft potentials at longitudes around $\pm 90^\circ$, i.e.\ above the neck region of the nucleus \citep{Preusker2015}, as observed also by \citet{Odelstad2015} for the period September 2014 - March 2015. The trend of generally decreasing $V_\textrm{s/c}$ peaked in late July, close to perihelion at about 1.25~AU, when there was a fairly abrupt transition to even more negative $V_\textrm{s/c}$ as the spacecraft concurrently went into the southern hemisphere. This was followed by an extended period of very negative $V_\textrm{s/c}$ that lasted well into September, with spacecraft potentials of -15~V to -25~V in the southern hemisphere and somewhat more modest -5~V to -10~V in the northern hemisphere, with occasional periods of $V_\textrm{s/c}$ close to zero (though persistently negative). Some tendency towards a longitude-modulation of the signal can be discerned in the figure in the northern hemisphere in mid-August, but the next time the spacecraft comes into the northern hemisphere, in mid-September, that modulation appears to be effectively gone.

In late September - early October, ($\sim$1.4~AU), Rosetta undertook a dayside excursion, leaving the habitual near-terminator (phase angle $\sim$90$^\circ$) orbits and heading out to a distance of almost 1500 km from the nucleus at a phase angle of $50^\circ$. During this excursion, the spacecraft potential was typically a few volts positive.

The period from November 2015 to northward equinox on March 21 2016 (1.5 - 2.7~AU) was characterised by steadily decreasing spacecraft cometocentric distance and increasing heliocentric distance. The spacecraft potential was consistently negative throughout this time period. From early November 2015 until late January 2016, $V_\textrm{s/c}$ showed a clear latitudinal dependence, with more negative $V_\textrm{s/c}$ above the southern (summer) hemisphere than the northern (winter) one. Before January, the aforementioned longitude-modulation was present in the northern hemisphere, though it decreased successively, being almost completely gone by mid-January. There was also a general trend towards more positive spacecraft potentials until mid-January, when, at a heliocentric distance of $\sim$2.2 AU, $V_\textrm{s/c}$ went more negative again, coinciding with the cometocentric distance going below about 80 km and the spacecraft going into the southern hemisphere. The following period until northward equinox was characterised by very negative $V_\textrm{s/c}$, -15~V to -20~V in the southern hemisphere and -5~V to -10~V, and little or no longitude modulation, in the northern hemisphere.

Around the time of northward equinox in late March 2016, Rosetta went on a second excursion, this time into the nightside of the coma, during which the spacecraft went out to $\sim$1000 km and a maximum phase angle of $\sim$160$^\circ$. During this excursion, the spacecraft potential went positive, up to about +5~V, at distances beyond $\sim$100 km of the nucleus.

Coming back to cometocentric distances below $\sim$ 30 km in the second half of April 2016 ($\sim$2.8~AU), the spacecraft potential again became negative, at generally about the same values as before the excursion (-15~V to -20~V in the southern hemisphere and -5~V to -10~V in the northern hemisphere). There was a slight further decrease in $V_\textrm{s/c}$ as the cometocentric distance dropped to to below 10 km in the second half of May 2016 ($\sim$3 AU), followed by more positive potentials at larger distances from the nucleus in June and the first half of July 2016 ($\sim$3.1 - 3.4~AU). It is worth noting that the dips in $V_\textrm{s/c}$ still occurred in the southern hemisphere, in spite of it being the winter hemisphere since northward equinox in late March 2016. This behaviour appeared to persist until EOM, although starting in late June 2016 the orbital configuration of the spacecraft was such that periapsis always occurred in the southern hemisphere, so from then on seasonal and radial effects could not be disentangled.

During August and September 2016, LAP1 was generally run in floating potential mode, only running sweeps in short intermittent intervals. Thus, for these last months of the mission, LAP1 floating potential measurements (with reversed sign) are plotted in Figure \ref{fig:lon_plot} (without any special annotations or identifying markings) along with the few scattered sweep measurements. These $-V_\textrm{F}$ values have been shifted by +4.5~V to account for their offset w.r.t.\ $-V_\textrm{ph}$ (c.f.\ Section \ref{sec:double_probes}), which is generally found to give good agreement with $-V_\textrm{ph}$ at the mode transitions throughout this time period.

\section{Summary and Conclusions}

We have presented RPC-LAP measurements of the spacecraft potential throughout Rosetta's stay at comet 67P. $V_{\textrm{s/c}}$ is mostly negative during this time, often below -10~V and sometimes below -20~V. We attribute this to a warm ($\sim$ 5 -- 10~eV) population of coma photo\-electrons that are present because the neutral gas is insufficiently dense for them to be effectively cooled by collisions with neutrals. Positive spacecraft potentials ($\sim$ 0 -- 5~V) were only observed in regions far from the nucleus, or above the more inactive areas on it, where the electron density is very low ($\lesssim$10 cm$^{-3}$) and where significant electron cooling by neutrals is not possible. Thus we conclude that the thermal flux of electrons in the cometary plasma, at the position of the spacecraft, is dominated by these warm, uncooled electrons throughout Rosetta's stay at the comet, notably also at and around perihelion where strongly negative spacecraft potentials were observed. The prevalence of these warm electrons correlate well with diurnal and seasonal variations that drive the neutral outgassing \citep{Hansen2016, Odelstad2015}, consistent with these electrons being predominantly produced at or inside the position of the spacecraft.

The pervasiveness of the warm electron population at the position of Rosetta around 67P is an effect of the comparatively low activity of the comet. Using neutral gas density measurements by ROSINA/COPS, \citet{Mandt2016} showed that Rosetta likely stayed outside the electron exobase, where the collision mean free path equals the local scale height of the neutral gas (which for an expanding cometary atmosphere is equal to the distance to the nucleus centre), during all of the mission. Nevertheless, \citet{Eriksson2017a} presented observations of cold (around or below 0.1~eV) electrons, interpreted as having undergone cooling close to the nucleus and propagated outward without any significant heating. The most obvious signature of the cold electrons were found to be intermittent, showing up in the data as pulses of typical duration between a few seconds and a few minutes. This is interpreted as filamentation of a cold plasma close to the nucleus, as observed in e.g.\ the hybrid simulations by \citet{Koenders2015}. A similar picture has been presented by \citet{Henri2017} for the intermittent appearance of a magnetic field-free plasma at Rosetta, interpreted as filaments or bubbles rising from an inner diamagnetic cavity.

It should be noted that while cold electrons thus existed around 67P during parts of the Rosetta mission, our results indicate that they never dominate the electron flux at the spacecraft position, at least for any extended times. This does not preclude that they may still sometimes dominate the electron density, as also inferred by \citet{Eriksson2017a} for some events, since a warm (10~eV) electron population would still contribute a flux ten times as high as a cold (0.1~eV) population at equal density. The statistical nature of our present study also cannot rule out the existence of some brief event of low spacecraft potential hiding in the dataset, which would indicate the near-absence of warm electrons. However, it is clear that such events must be rare exceptions.

The ubiquitous presence of warm electrons at 67P implies that models assuming cold electrons, presumably applicable to higher-activity comets like 1P/Halley at its spacecraft encounters in 1986 (c.f.\ \citet{Häberli1996} and references therein), are not valid for 67P at Rosetta's position (nor, presumably, outside of it). Since the electron pressure is about two orders of magnitude higher if there is no electron cooling, the plasma beta parameter (ratio of kinetic to magnetic pressure, a fundamental parameter for determining plasma dynamics) will exceed unity already at plasma densities of about 25~cm$^{-3}$ for $T_\textrm{e} = 10$ eV, compared to 2,500 cm$^{-3}$ for 0.1 eV. A large region of high-beta plasma will have implications for the stability and dynamics of the cometary plasma. This may be the reason for the wealth of dynamics seen in the particle-in-cell simulations of the plasma environment of 67P by \citet{Deca2017}. Another example is the lower hybrid drift waves investigated by \citet{Karlsson2017} and \citet{Andre2017}.

We have shown, both theoretically and empirically, that the two LAP probes attain the same floating potentials in the spacecraft potential well in the absence of any ambient electric field, confirming and explaining the aptitude of the double-probe floating potential technique for measuring electric fields in the denser parts of the coma, used e.g.\ by \citet{Karlsson2017}. We have also shown that the LAP1 floating potential measurements pick up the same fraction of $V_{\textrm{s/c}}$ as does the photoelectron knee potential in the Langmuir probe sweeps.

We have combined measurements by RPC-ICA and RPC-LAP to determine what fraction of the spacecraft potential is observed by LAP and the ICA energy offset. We find numerous cases of good correspondence between the two instruments, increasing our confidence in the their accuracy. The correlation disappears intermittently, coincident with weakened ion fluxes. This is interpreted as temporary loss of local ionisation, at least within the ICA field of view, causing the ion flux into the instrument to be dominated by accelerated ions, the energy of which is not strongly dependent on $V_{\textrm{s/c}}$. Measurement noise is also found to drown out the correlation during quiet periods when $V_{\textrm{s/c}}$ changes very little. The fraction of $V_{\textrm{s/c}}$ picked up by LAP1 is found to vary between about 0.7 and 1, indicating that a correction factor between about 1 and 1.4 should be applied to the LAP1 measurements to obtain the full $V_{\textrm{s/c}}$. The ICA energy offset is estimated to 13.7 eV, with a 95\% confidence interval between 12.5 and 15.0 eV.

We have also investigated the possible contribution of ambient positive ions to the spacecraft potential, finding it to be generally negligible throughout the mission. This gives increased confidence to previously published electron density and temperature estimates derived from the spacecraft potential assuming a current balance to the spacecraft of only ambient plasma electrons and spacecraft photoelectrons \citep{Galand2016, Heritier2017a, Harja2017}. The possible impact of supra\-thermal electrons on $V_{\textrm{s/c}}$ could potentially be determined with the aid of the Ion and Electron Sensor (RPC-IES) onboard Rosetta, but such an investigation is deferred to a future paper.

\section*{Acknowledgements}

Rosetta is a European Space Agency (ESA) mission with contributions from its member states and the National Aeronautics and Space Administration (NASA). The work on RPC-LAP and RPC-ICA data was funded by the Swedish National Space Board under contracts 108/12, 109/12, 149/12 and 168/15. This work has made use of the AMDA and RPC Quicklook database, provided by a collaboration between the Centre de Donn{\'e}es de la Physique des Plasmas (CDPP) (supported by CNRS, CNES, Observatoire de Paris and Universit{\'e} Paul Sabatier, Toulouse) and Imperial College London (supported by the UK Science and Technology Facilities Council). The data used in this paper will soon be made available on the ESA Planetary Science Archive and is available upon request until that time.




\bibliographystyle{mnras}
\bibliography{references.bib} 



%
%


\bsp	
\label{lastpage}
\end{document}